\documentclass[aps,pra,reprint,superscriptaddress,nofootinbib,floatfix]{revtex4-2}

\usepackage[T1]{fontenc}
\usepackage[utf8]{inputenc}
\usepackage{lmodern}
\usepackage{microtype}
\usepackage{graphicx}
\usepackage{amsmath,amssymb,bm}
\usepackage{booktabs}
\usepackage{array}
\usepackage{placeins}
\usepackage[hidelinks]{hyperref}

\graphicspath{{figures/}}
\newcolumntype{L}[1]{>{\raggedright\arraybackslash}p{#1}}

\begin{document}

\title{Nonreciprocal Control of Spatiotemporal Floquet Dynamics in Levitated Optomechanical Arrays}

\author{Guangyu Zhang}
\affiliation{Center for Quantum Technology Research and Key Laboratory of Advanced Optoelectronic Quantum Architecture and Measurements (MOE), School of Physics, Beijing Institute of Technology, Beijing 100081, China}
\affiliation{Beijing Key Laboratory of Quantum Matter State Control and Ultra-Precision Measurement Technology, School of Physics, Beijing Institute of Technology, Beijing 100081, China}
\author{Qihao Guo}
\affiliation{Department of Applied Physics, Xi'an Jiaotong University, Xi'an, Shaanxi 710049, China}
\author{Zhang-qi Yin}
\email{zqyin@bit.edu.cn}
\affiliation{Center for Quantum Technology Research and Key Laboratory of Advanced Optoelectronic Quantum Architecture and Measurements (MOE), School of Physics, Beijing Institute of Technology, Beijing 100081, China}
\affiliation{Beijing Key Laboratory of Quantum Matter State Control and Ultra-Precision Measurement Technology, School of Physics, Beijing Institute of Technology, Beijing 100081, China}

\date{October 15, 2026}

\begin{abstract}
Periodically driven photonic and optomechanical systems provide a versatile
setting for realizing collective Floquet dynamics, yet the relation between
temporal locking and continuous rotation of the finite-$q$ pattern remains poorly understood. We
show that a nonreciprocal ring of parametrically driven optically levitated
particles allows the spatial phase to be controlled while collective
temporal locking is preserved. Nonreciprocity polarizes the counterpropagating finite-wave-vector modes and drives a transition from the pinned state to persistent directional spatial-phase drift. The normalized spatial-phase drift velocity is independently encoded in the splitting of the two dominant spectral branches, and its direction reverses with the coupling asymmetry. The pinned regime exhibits a laboratory frame period-two time-crystalline response, whereas the rotating regime forms a distinct temporally locked spatiotemporally ordered state. The same correspondence between spatial-phase drift and spectral splitting persists down to few-particle rings, indicating that the effect does not require a large array. These results establish nonreciprocity as a route to independently control spatial-phase drift and temporal synchronization in driven levitated optomechanical systems.

\end{abstract}

\maketitle

\section{Introduction}

Temporal modulation provides a powerful route for controlling collective
dynamics in photonic and optomechanical systems. Periodic driving can
reshape mode structure and generate Floquet states with nontrivial spatial
and temporal organization
\cite{asgari2024photonic,redondo2024nonreciprocal}.
In parametrically driven systems, finite-wave-vector responses commonly
involve standing or counterpropagating modes
\cite{zhang1997pattern,martel2000counterpropagating,martel2003dispersion}.
Optically levitated particles offer a particularly flexible platform
because their nonlinear motion can be driven and resolved at the
single-particle level
\cite{gieseler2014nonlinear,ricci2017stochastic,millen2020optomechanics},
while programmable traps, particle arrays, and mediated interactions
enable increasingly controlled collective dynamics
\cite{yan2023ondemand,vijayan2023scalable,siegel2025arrays,
vijayan2024cavity}. These developments make levitated arrays a natural
setting for studying finite-wave-vector Floquet states with simultaneously
accessible temporal and spatial dynamics.

Robust temporal order and persistent spatial motion have both been widely
explored in driven systems, although largely along separate lines.
Discrete time crystals provide a prominent example of subharmonic temporal
responses that remain locked over finite parameter ranges
\cite{khemani2016phase,yao2017discrete,zhang2017observation,
choi2017observation}. Recent work has extended time-crystalline behavior
to dissipative parametric oscillators, Rydberg gases, nonlinear optical
cavities, spin systems, actively mode-locked lasers, sensing protocols,
and higher-dimensional settings
\cite{yithomas2024theory,wu2024rydberg,li2024cavity,
greilich2024continuous,weng2026laser,wahl2024topological,
liu2025bifurcation,moon2025rondeau,moon2026sensing,
switzer2026twodimensional}. In parallel, space-time crystals, traveling
waves, and drifting parametrically excited patterns demonstrate that
spatial motion can persist under periodic forcing
\cite{smits2018spacetime,xu2018spacetime,zhao2026discrete,
marin2023drifting,guan2023traveling,kucher2025oscillons}.
Sliding and depinning phenomena in modulated media provide related examples
in which the phase of a spatial pattern acts as a collective coordinate
\cite{gruner1988dynamics,lee1974conductivity,lee1979depinning,
floria1996frenkel}.

A finite-wave-vector Floquet state naturally brings these two aspects
together. For a state formed by counterpropagating components, one
combination of their phases determines the collective phase relative to
the drive, while another determines the spatial phase of the modulation
\cite{cross1993pattern,martel2000counterpropagating,
martel2003dispersion}. Temporal locking therefore need not, in principle,
fix the spatial phase. The unresolved question is whether these two
collective coordinates can remain dynamically distinct within the same
Floquet state, so that temporal synchronization survives while the spatial
pattern acquires persistent directed motion.

Nonreciprocal interactions provide a natural mechanism for separating
these roles. Direction-dependent couplings break the equivalence between
counterpropagating modes and can generate directional pattern dynamics,
traveling states, and nonequilibrium phase transitions
\cite{fruchart2021nonreciprocal,brauns2024nonreciprocal,
mason2025dynamical}. Nonreciprocity has also been linked to space-time
ordering, synchronization, and time-crystalline dynamics
\cite{raskatla2024continuous,nadolny2025nonreciprocal,
morrell2026nonreciprocal}. In levitated-particle systems, tunable optical
coupling and controllable non-Hermitian and nonreciprocal dynamics have
already been demonstrated
\cite{rieser2022tunable,reisenbauer2024nonhermitian,
rudolph2024quantum}. Nonreciprocity therefore offers a direct route to bias
the spatial component of a driven pattern without changing the period of
the external modulation.

\begin{figure*}[!t]
\centering
\includegraphics[width=0.8\textwidth]{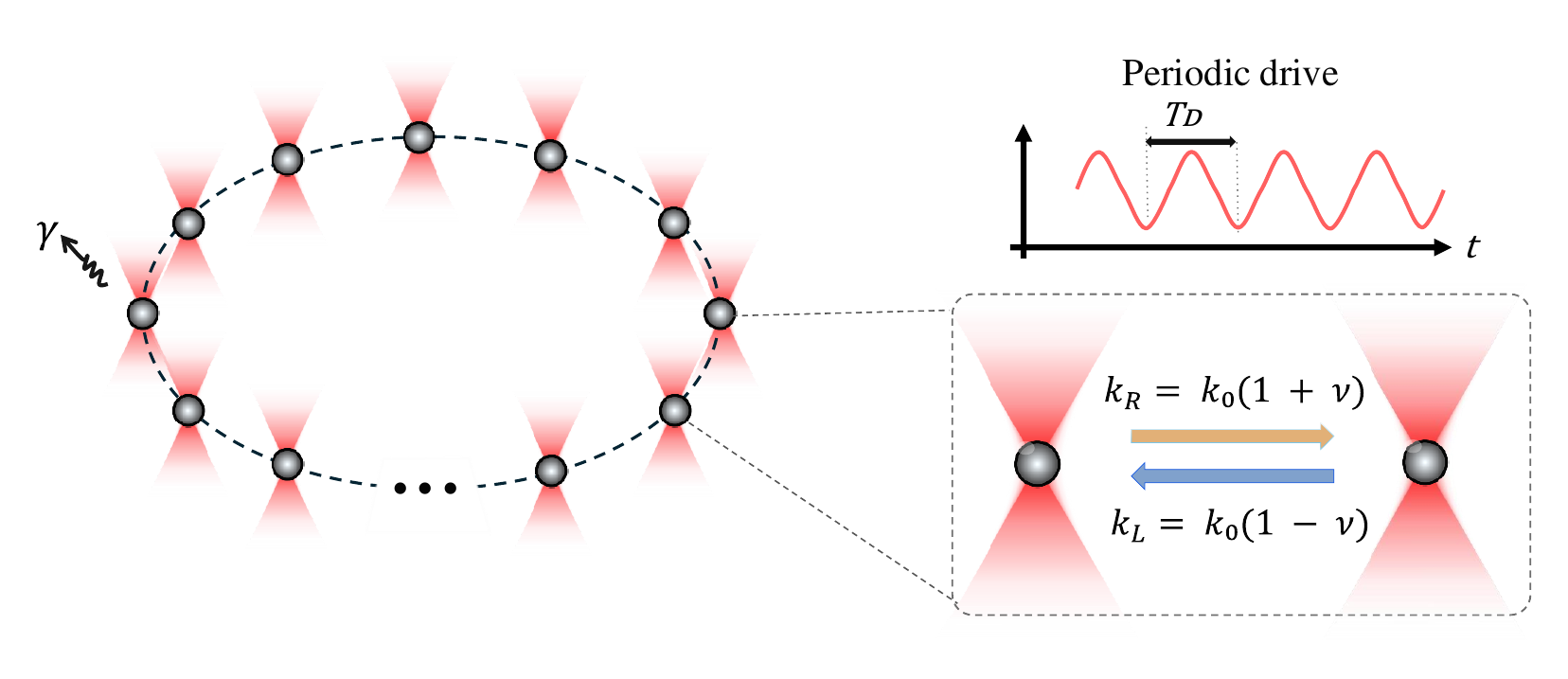}
\caption{\textbf{Driven nonreciprocal oscillator ring.} Schematic of a periodic
ring of $N$ optically levitated particles confined in individual optical traps. Each particle is parametrically modulated with period $T_D=2\pi/\Omega_D$ and damped at rate $\gamma$. The directed nearest-neighbor couplings are $k_R=k_0(1+\nu)$ and $k_L=k_0(1-\nu)$.}
\label{fig:system}
\end{figure*}

Here we study this mechanism in a parametrically driven ring of optically
levitated particles with direction-dependent nearest-neighbor interactions.
We show that nonreciprocity polarizes the counterpropagating
finite-wave-vector pair and unlocks the spatial phase into persistent
directional motion while the collective temporal phase remains locked.
The phase drift velocity is encoded in the splitting of the two dominant
spectral branches and reverses with the coupling asymmetry. The pinned
regime retains a laboratory frame period-two time-crystalline response,
whereas the rotating regime forms a distinct temporally locked
spatiotemporally ordered state. The same phase--spectral relation
persists in few-particle rings. These results identify nonreciprocity as a
route to reversible control of pattern rotation within temporally
synchronized Floquet states.

\section{Model and Floquet Response}

\subsection{Model}

We consider a periodic ring of $N$ optically levitated particles as shown in Figure~\ref{fig:system}, each confined in an individual
optical trap and subject to periodic parametric modulation. 
Levitated particles
provide well-controlled nonlinear mechanical oscillators, for which
parametric driving and nonlinear dynamics have been demonstrated
experimentally
\cite{gieseler2014nonlinear,ricci2017stochastic,millen2020optomechanics}.
The displacement $x_i$ of the particle $i$ obeys
\begin{align}
\ddot{x}_i+\gamma\dot{x}_i={}&
-\omega_0^2\left[1+A_D\cos(\Omega_D t)\right]x_i e^{-2x_i^2}
-\beta x_i^3 \nonumber\\
&+k_R(x_{i-1}-x_i)+k_L(x_{i+1}-x_i)+\xi_i(t),
\label{eq:model}
\end{align}
with periodic boundary condition $x_{i+N}=x_i$. Here $\omega_0$ is the natural frequency of particle, $\gamma$ is the damping rate, $A_D$ and $\Omega_D$ are the amplitude and angular frequency of the periodic drive. The coefficient $\beta$ controls the additional cubic onsite nonlinearity. The directed nearest-neighbor couplings are
\begin{equation}
k_R=k_0(1+\nu),\qquad k_L=k_0(1-\nu),
\label{eq:directed-coupling}
\end{equation}
where $k_0$ is the mean coupling and $\nu$ quantifies nonreciprocity. Reciprocal coupling is recovered at $\nu=0$, and changing the sign of $\nu$ interchanges the left- and right-directed couplings. 
Such tunable optical interactions 
have recently been realized between optically levitated
particles\cite{rieser2022tunable,reisenbauer2024nonhermitian}.

Thermal fluctuations are represented by independent Gaussian white noise with zero mean, $\xi_i(t)$ satisfies
\begin{equation}
\langle\xi_i(t)\xi_j(t')\rangle=2\gamma T\delta_{ij}\delta(t-t').
\label{eq:noise}
\end{equation}
where $T$ is the dimensionless temperature relative to the characteristic energy scale of the optical trap. Unless stated otherwise, we use $N=100$, $\omega_0=1$, $\gamma=0.15$, $\beta=1$, $k_0=0.1$, and $\Omega_D=2.07$. The model construction, nondimensionalization, and an illustrative
mapping to optical-tweezer parameters are given in Secs.~S1 and S6 of
the Supporting Information.

\subsection{Linear Floquet response and nonreciprocal polarization}

\begin{figure}[!t]
\centering
\includegraphics[width=\columnwidth]{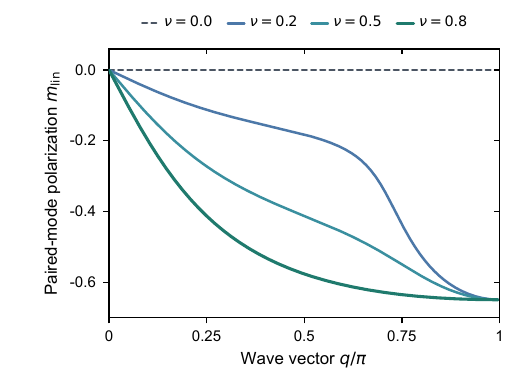}
\caption{\textbf{Nonreciprocal polarization of the finite-wave-vector
Floquet pair.} 
Linear polarization $m_{\rm lin}$ of the dominant paired eigenmode as a
function of wave vector for several values of the nonreciprocity $\nu$.}
\label{fig:linear-directionality}
\end{figure}

Before addressing the nonlinear states, we first identify how
nonreciprocity modifies the finite-wave-vector response at the linear
level. Linearizing Eq.~\eqref{eq:model} around $x_i=0$, the directed
nearest-neighbor coupling of a Fourier component with wave vector $q$
contributes
\begin{equation}
k_R(e^{-iq}-1)+k_L(e^{iq}-1)
=
2k_0(\cos q-1)-2ik_0\nu\sin q.
\label{eq:q-coupling}
\end{equation}
The reciprocal contribution is even in $q$, whereas the nonreciprocal
contribution is odd. Nonreciprocity therefore introduces a directional
bias between counterpropagating components.

Near the principal parametric resonance, we remove the fast oscillation
at $\Omega_D/2$ and describe the dynamics in terms of slowly varying
complex amplitudes, following the standard rotating-wave treatment of
parametrically driven oscillators and waves
\cite{dykman1998phaseflip,martel2000counterpropagating,
yithomas2024theory}.
We denote the Fourier components of the resulting slow envelope by
$\psi_q$. Parametric driving couples $\psi_q$ to
$\psi_{-q}^{*}$. With
$\bm{\Psi}_q=(\psi_q,\psi_{-q}^{*})^{\mathsf T}$, their near-resonant
dynamics is described by
\begin{align}
\partial_t\bm{\Psi}_q&=
\begin{pmatrix}
-\gamma/2-g_q+i\delta_q & p\\
p & -\gamma/2+g_q-i\delta_q
\end{pmatrix}
\bm{\Psi}_q,
\label{eq:paired-linear}
\end{align}
where
$p=A_D\omega_0^2/(4\omega)$,
$g_q=k_0\nu\sin q/\omega$, and
$\delta_q=[\omega_0^2-\omega^2+2k_0(1-\cos q)]/(2\omega)$.
The nonreciprocal term $g_q$ enters the two components with opposite
signs and therefore changes their relative weight in the leading paired
eigenmode.

If $(u_q,v_q)^{\mathsf T}$ denotes the corresponding right eigenvector,
we quantify this imbalance by
\begin{equation}
m_{\rm lin}(q)=
\frac{|u_q|^2-|v_q|^2}
{|u_q|^2+|v_q|^2}.
\label{eq:linear-polarization}
\end{equation}
As shown in Figure~\ref{fig:linear-directionality}, the reciprocal system
has a balanced pair, whereas nonreciprocity produces a finite directional
polarization. Thus, the primary linear effect of nonreciprocity is to
polarize the internal composition of the counterpropagating Floquet pair.

The full linear Floquet analysis, the paired-RWA derivation and onset,
and the detailed eigenvector analysis are given in Sec.~S2 of the
Supporting Information. Whether this linear directional bias develops into
a pinned or rotating state is determined by the nonlinear saturated
dynamics, which motivates the collective-phase description introduced
next.

\subsection{Collective phases of the finite-wave-vector state}

In the nonlinear saturated state, the same slow envelope $\psi_i$ is
dominated by the pair selected at $\pm q_0$. Such decompositions into
counterpropagating complex amplitudes are standard in parametrically
driven pattern-forming systems
\cite{zhang1997pattern,martel2000counterpropagating}.
We therefore write
\begin{equation}
\psi_i\simeq
B_+e^{iq_0 i}+B_-e^{-iq_0 i},
\qquad
B_\pm=R_\pm e^{-i\phi_\pm},
\label{eq:two-branch}
\end{equation}
where $q_0>0$ is determined from the late-time spatial spectrum,
$R_\pm$ are the amplitudes of the selected counterpropagating branches,
and $\phi_\pm$ are their phases. We then define the collective temporal
phase $\Sigma$ and collective phase difference $\Delta$ that determines
the spatial phase by
\begin{equation}
\Sigma=\phi_++\phi_-,
\qquad
\Delta=\phi_+-\phi_-.
\label{eq:collective-phases}
\end{equation}

The two combinations have distinct physical meanings. A common phase
shift of the counterpropagating branches changes the phase of the response
relative to the periodic drive and is therefore encoded in $\Sigma$.
By contrast, a spatial translation changes the two branch phases with
opposite signs and is therefore encoded in $\Delta$. For balanced
branches, $R_+=R_-=R$, Eq.~\eqref{eq:two-branch} reduces to
\begin{equation}
\psi_i=
2R e^{-i\Sigma/2}
\cos\left(q_0i-\frac{\Delta}{2}\right),
\label{eq:phase-geometry}
\end{equation}
which makes this separation explicit: $\Sigma$ controls the temporal
phase, whereas $\Delta/2$ is the spatial phase of the finite-$q$ pattern
and fixes its position, or spatial registry, relative to the particle array.
We therefore use $\Delta$ as the spatial-phase coordinate. For weak branch imbalance, Eq.~\eqref{eq:phase-geometry}
remains a useful geometric interpretation of the two collective phases.

\begin{figure*}[!t]
\centering
\includegraphics[width=0.8\textwidth]{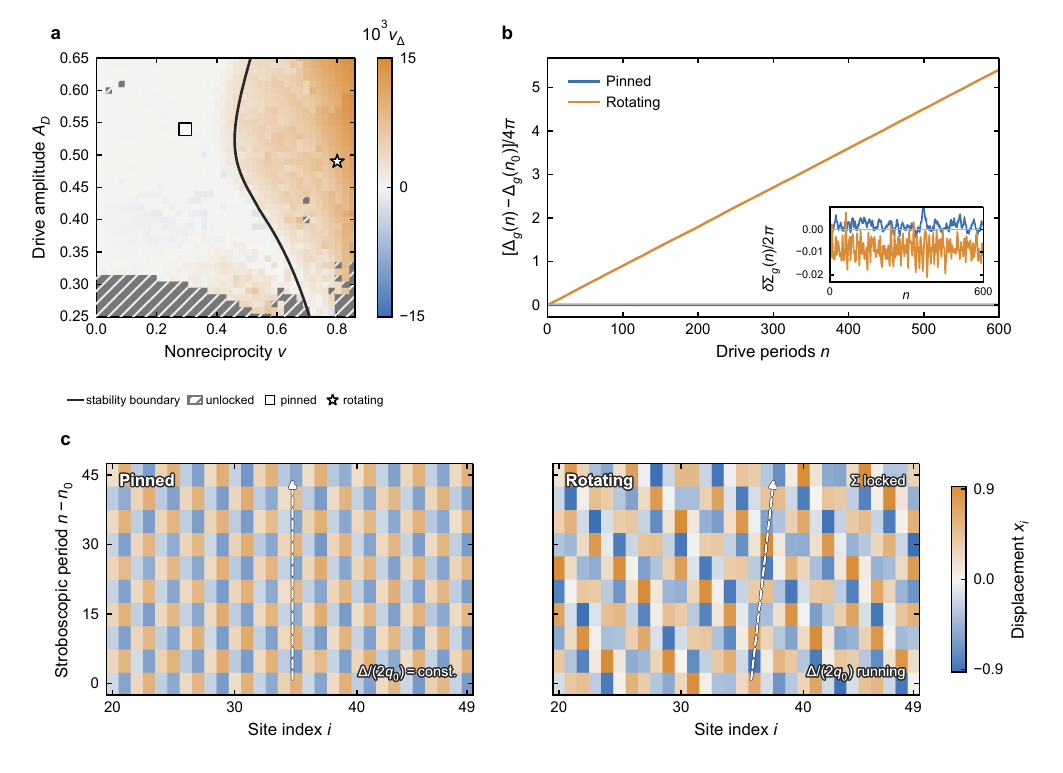}
\caption{\textbf{Temporal locking in pinned and rotating states.} \textbf{a}, Signed normalized spatial-phase drift velocity $10^3v_\Delta$ in the nonreciprocity--drive-amplitude plane, using five independent realizations per grid point and displaying the median where at least three pass the selection criteria. The solid curve is the nonlinear pinned-state stability boundary. Hatching marks grid points without the required temporal-locking seed consensus. The square and star mark the pinned $(\nu,A_D)=(0.295,0.54)$ and rotating $(0.80,0.49)$ representatives. \textbf{b}, Accumulated pattern displacement $[\Delta_g(n)-\Delta_g(n_0)]/(4\pi)$ for one trajectory of each representative. The inset shows the temporal-phase change $\delta\Sigma_g(n)/(2\pi)$, where $\delta\Sigma_g(n)=\Sigma_g(n)-\Sigma_g(n_0)$. \textbf{c}, laboratory frame stroboscopic displacements $x_i(nT_D)$ over a 30-site window of the same trajectories.}
\label{fig:depinning}
\end{figure*}

In the full numerical dynamics, we use the corresponding global phases
$\Sigma_g$ and $\Delta_g$ extracted from the spatially coherent branch
fields. These quantities allow temporal locking and spatial-phase drift to be
diagnosed independently and form the primary variables used in
Sec.~III.
The nonlinear two-branch reduction and the resulting amplitude and
phase equations are derived in Sec.~S4 of the Supporting Information.
We select $q_0$ from the late-time spatial spectrum and extract
$\Sigma_g$ and $\Delta_g$ from the corresponding branch fields, as
specified in Sec.~S3. The spatial-phase drift rate
$\mathrm d\Delta_g/\mathrm dn$ defines the normalized spatial-phase drift velocity
\begin{equation}
v_\Delta=
\frac{1}{4\pi}\frac{d\Delta_g}{dn},
\label{eq:registry-velocity}
\end{equation}
measured in pattern wavelengths per drive cycle $n=t/T_D$. The
corresponding real-space displacement rate is $(2\pi/q_0)v_\Delta$
lattice sites per drive cycle. Hereafter we refer to $v_\Delta$ simply as
the phase drift velocity. The numerical protocol and observable
extraction are detailed in Sec.~S3.

The full equations of motion are integrated on a periodic ring with an
exponential-damping Langevin update. Unless specified otherwise, the
simulations use $N=100$, 40 time steps per drive cycle, and the model
parameters given above. Initial transients are discarded before the
late-time spatial mode, collective phases, phase drift velocity, and
laboratory frame spectrum are evaluated. Independent initial-condition
and noise seeds are used where multiple realizations are reported.
Figure legends state the number of realizations and the displayed
summary statistics; integration, selection, and validation details are
provided in Sec.~S3 of the Supporting Information.

\section{Results}

\subsection{Temporal locking and spatial-phase drift}

We first ask whether sustained spatial-phase drift can develop while the
temporal phase remains locked. We characterize temporal order
through collective phase locking
\cite{yao2017discrete,else2020discrete,yithomas2024theory}. 
For the parameter scan in
Figure~\ref{fig:depinning}a, we operationally classify a trajectory as
temporally locked when
$r_\Sigma \geq 0.97$,
$\left|\mathrm d\Sigma_g /\mathrm dn\right|
\leq 5\times10^{-3}$, 
where $n=t/T_D$ and $r_\Sigma$ denotes the late-time mean,
overlap-weighted circular coherence of the local temporal phases, with
$r_\Sigma=1$ corresponding to perfect across-site temporal-phase alignment. These thresholds are fixed operational tolerances for classifying finite-time trajectories,
rather than physical critical values. The construction and calibration of $r_\Sigma$, together with
threshold-sensitivity tests, are given in Sec.~S3 of the Supporting Information.

Within the temporally locked region, Figure~\ref{fig:depinning}a reveals
two distinct spatial-phase behaviors. A broad region has a pinned spatial
phase with $v_\Delta\simeq0$, while persistent spatial-phase drift with
finite signed $v_\Delta$ develops as
nonreciprocity is increased. The nonlinear stability boundary of the
pinned solution closely tracks the onset of this finite spatial-phase drift,
linking the loss of the stationary spatial phase to the emergence of the
rotating state. The two marked points provide representative
examples of these regimes.

\begin{figure*}[!t]
\centering
\includegraphics[width=0.8\textwidth]{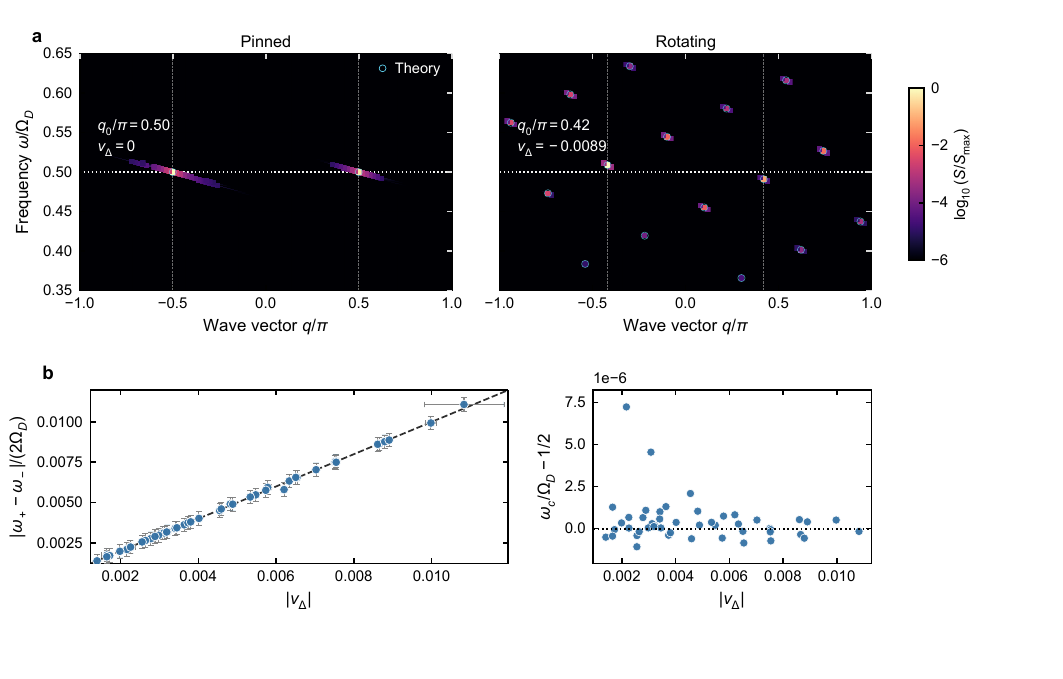}
\caption{\textbf{Momentum-frequency signature of pinned and rotating
states.}
\textbf{a}, Joint spectra
$\log_{10}[S(q,\omega)/S_{\max}]$ for the deterministic pinned and
rotating representatives (one trajectory each), obtained from an 8000-cycle observation
window using physical Fourier bins without zero padding or
interpolation. Open circles mark the peak positions predicted from the
phase-derived $v_\Delta$. The horizontal line marks
$\omega/\Omega_D=1/2$.
\textbf{b}, Left, normalized spectral half-splitting
$|\omega_+-\omega_-|/(2\Omega_D)$ versus the phase-derived
$|v_\Delta|$ for 48 accepted trajectories from a predefined 99-point parameter grid.
The dashed line denotes $y=x$. Error bars show the corresponding
phase-slope and Fourier-bin uncertainties. Right, spectral-center
deviation $\omega_c/\Omega_D-1/2$ versus $|v_\Delta|$, where
$\omega_c=(\omega_++\omega_-)/2$.}
\label{fig:spectrum}
\end{figure*}

The corresponding phase trajectories make the separation between the
two collective coordinates explicit. In Figure~\ref{fig:depinning}b, the
pinned state shows no net accumulation of $\Delta_g$, whereas the
rotating state exhibits an approximately linear increase of the
pattern displacement. By contrast, the inset shows no comparable
systematic drift of $\Sigma_g$ for either trajectory. Thus, the
transition from the pinned to the rotating state occurs without loss of
the collective temporal locking diagnosed by $\Sigma_g$.

This distinction is also visible directly in the microscopic dynamics.
Figure~\ref{fig:depinning}c shows the laboratory frame stroboscopic
displacements $x_i(nT_D)$ over a 30-site window. 
In the pinned state, the spatial phase remains fixed and the pattern
exhibits the period-two sign alternation characteristic of a discrete
time-crystalline order\cite{yao2017discrete,yithomas2024theory}. 
In the rotating state, 
the pattern rotates continuously between stroboscopic snapshots
while $\Sigma_g$ remains locked. Consequently, a fixed laboratory site
does not exhibit a strict period-two trace in the rotating state.
The rotating regime is therefore characterized by collective temporal
locking coexisting with continuous spatial-phase drift.
The pinned period-two response also remains locked at several sampled
drive detunings around the representative point, with negligible spatial-phase
drift; the finite-window test is given in Sec.~S5 of the Supporting Information.

\subsection{Spectral signatures of spatial-phase drift}

The preceding phase analysis showed that the temporal phase
can remain locked while the spatial phase acquires a persistent
drift. We now ask whether this separation has an independent
laboratory frame spectral signature. We compute the joint spectrum from
the measured displacements,
\begin{equation}
S(q,\omega)=
\left|\sum_{i,n}x_i(t_n)e^{-i(qi-\omega t_n)}\right|^2,
\label{eq:joint-spectrum}
\end{equation}
using the full laboratory frame trajectory. A rotating pattern changes the
relative phase velocities of the two counterpropagating components and
therefore predicts opposite frequency shifts about the half-drive
frequency\cite{marin2023drifting,kiselev2025symmetry}.

When $\mathrm d\Sigma_g/\mathrm dn\simeq0$, 
the two branch phases acquire opposite
drifts through $\Delta_g$. Using the definition of $v_\Delta$ in
Eq.~\eqref{eq:registry-velocity}, their frequencies satisfy
\begin{equation}
\frac{\omega_\pm}{\Omega_D}
=
\frac{1}{2}\pm v_\Delta ,
\label{eq:branch-shift}
\end{equation}
up to the sign convention used to label the two branches. Equivalently,
\begin{equation}
\frac{|\omega_+-\omega_-|}{2\Omega_D}
=
|v_\Delta|,
\qquad
\frac{\omega_++\omega_-}{2\Omega_D}
=
\frac{1}{2}.
\label{eq:spectral-relation}
\end{equation}
Spatial-phase drift should therefore split the two counterpropagating
branches symmetrically about the half-drive frequency without shifting
their mean temporal carrier.

This prediction is directly visible in Figure~\ref{fig:spectrum}a.
For the pinned state, the dominant peaks at both signs of $q_0$ remain
at $\omega/\Omega_D\simeq1/2$. In the rotating state, the two branches
instead appear at opposite frequency offsets about the same center.
The peak positions predicted independently from the phase-derived
$v_\Delta$ coincide with the observed spectral peaks. The laboratory frame
frequency splitting is therefore quantitatively tied to the rotating
spatial phase.

Figure~\ref{fig:spectrum}b tests this relation across the rotating
regime. The spectral half-splitting follows the independently measured
$|v_\Delta|$ along the predicted $y=x$ relation, while the spectral
center remains close to $\Omega_D/2$. Thus, the phase and spectral diagnostics identify the same collective
degree of freedom: the temporal center remains locked, whereas spatial-phase
drift appears as an antisymmetric frequency splitting of the
counterpropagating branches. Having established this phase--spectrum
correspondence, we next ask how nonreciprocity controls the direction
and magnitude of the spatial-phase drift.

\begin{figure}[!t]
\centering
\includegraphics[width=\columnwidth]{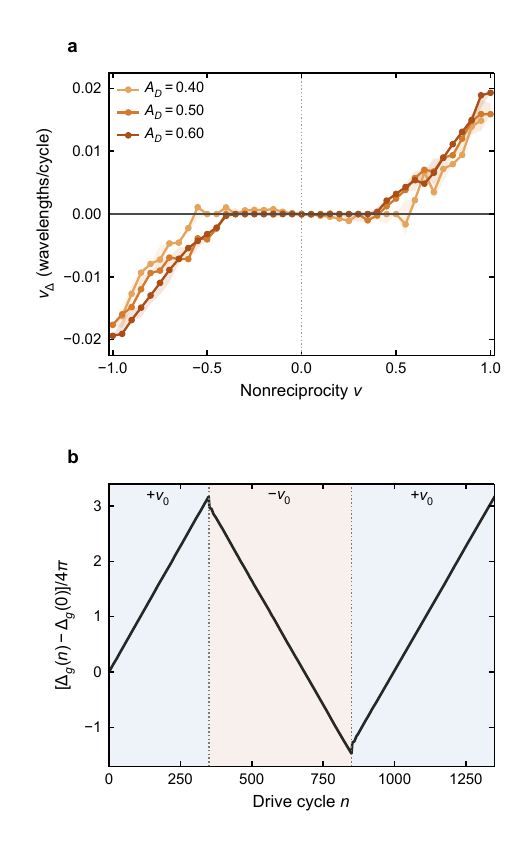}
\caption{\textbf{Nonreciprocal control of spatial-phase drift.}
\textbf{a}, Signed phase drift velocity $v_\Delta$, in pattern wavelengths
per drive cycle, as a function of $\nu$ for $A_D=0.40$, $0.50$, and
$0.60$. Five independent realizations are simulated per grid point. Points denote the median over temporally locked trajectories,
and bands show the interquartile range.
\textbf{b}, Accumulated pattern displacement under the switching
protocol $+\nu_0\rightarrow-\nu_0\rightarrow+\nu_0$, with
$\nu_0=0.80$, $A_D=0.49$, and $T=10^{-3}$ in one trajectory. Dotted lines mark the
switching cycles.}
\label{fig:control}
\end{figure}

\subsection{Nonreciprocal control of spatial-phase dynamics}

Reversing the sign of $\nu$ interchanges
$k_R$ and $k_L$ and therefore reverses the directional asymmetry of the
coupling. Spatial-phase drift generated by this asymmetry is consequently
expected to reverse its direction under $\nu\rightarrow-\nu$. 
Directional collective motion induced by nonreciprocal interactions has
also appeared in other active, oscillator, and optically coupled systems
\cite{fruchart2021nonreciprocal,raskatla2024continuous,
reisenbauer2024nonhermitian}. Here, we test this control directly through the phase drift velocity.

Figure~\ref{fig:control}a shows this behavior over three drive
amplitudes. The phase drift velocity remains small near the reciprocal
limit and develops opposite signs for positive and negative
nonreciprocity, approximately satisfying
$v_\Delta(-\nu)=-v_\Delta(\nu)$. Its magnitude also varies continuously
with $\nu$ and depends on the drive amplitude, showing that
nonreciprocity controls the direction of spatial-phase drift while tuning its speed
within the accessible nonlinear states.

The directional control persists when the coupling asymmetry is reversed
during a single trajectory. In Figure~\ref{fig:control}b, switching
$+\nu_0\rightarrow-\nu_0\rightarrow+\nu_0$ reverses the slope of the
accumulated spatial-phase change and restores its original sign after
the second switch. Linear fits to the three settled intervals give
$0.00911$, $-0.00900$, and $0.00898$ pattern wavelengths per drive cycle,
respectively. The recovery of both sign and magnitude shows that the
direction of spatial-phase drift follows the imposed coupling asymmetry rather than
being fixed solely by the initial state. Nonreciprocity therefore
provides reversible control of the spatial-phase dynamics.

The reversal tests identify nonreciprocity as a direct control parameter
for the spatial-phase drift. We next examine whether the relation between
spatial-phase dynamics and its spectral signature remains stable under
perturbations and changes in system size.

\subsection{Robustness of spatial-phase dynamics}

\begin{figure}[!t]
\centering
\includegraphics[width=\columnwidth]{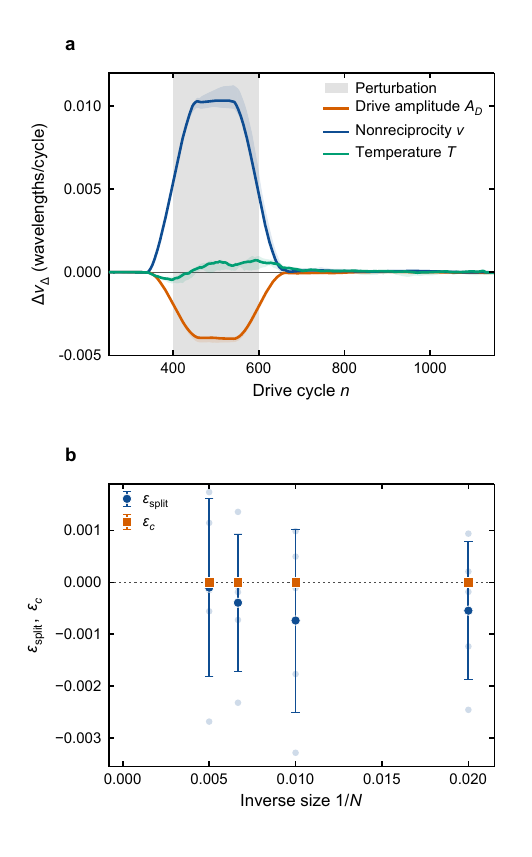}
\caption{\textbf{Recovery and finite-size persistence of the
phase--spectral relation.}
\textbf{a}, Change in phase drift velocity during finite pulses of the
drive amplitude $A_D$, nonreciprocity $\nu$, and effective temperature
$T$. The shaded interval marks the perturbation. Curves show medians
over 20 realizations and bands show interquartile ranges.
\textbf{b}, Finite-size residuals for $N=50$, $100$, $150$, and $200$, with five independent realizations per size.
The splitting residual is
$\epsilon_{\rm split}
=[\delta\omega_{\rm spec}-|v_\Delta|]/\delta\omega_{\rm spec}$,
where
$\delta\omega_{\rm spec}=|\omega_+-\omega_-|/(2\Omega_D)$,
and the center-locking residual is
$\epsilon_c=\omega_c/\Omega_D-1/2$.
Faint points denote individual realizations; large symbols and error
bars show the corresponding means and standard deviations.}
\label{fig:robustness}
\end{figure}

We next test whether the spatial-phase dynamics and their associated spectral
relation remain stable beyond an unperturbed representative trajectory.
Figure~\ref{fig:robustness}a applies finite pulses to the drive amplitude,
nonreciprocity, and noise strength. Each perturbation temporarily changes
the phase drift velocity, but after the pulse the velocity returns close to
its pre-pulse value. The recovery across different perturbation channels
shows that the spatial-phase drift is a dynamically restored property of the
driven state rather than a transient response to a particular parameter
history.

We then test whether the phase--spectrum correspondence persists as the
system size is varied. Changing the ring size modifies the discrete set
of allowed wave vectors, $q_m=2\pi m/N$, so the selected $q_0$ need not
remain identical across different $N$. We therefore compare quantities
that test the relation itself rather than the absolute phase drift velocity.
As shown in Figure~\ref{fig:robustness}b, both the splitting residual
$\epsilon_{\rm split}$ and the center-locking residual $\epsilon_c$ remain
close to zero for $N=50$, $100$, $150$, and $200$. Within the tested
range, finite-size changes therefore modify the available spatial modes
without producing a resolved breakdown of either the phase--spectral
relation or the locking of the spectral center to $\Omega_D/2$.

Taken together, these tests show that the rotating dynamics are not
identified only at a single parameter point or system size. The phase drift
velocity recovers after finite perturbations, while its quantitative
relation to the momentum-frequency splitting persists across the tested
ring sizes. This provides the robustness evidence for the
spatiotemporal dynamics characterized in the preceding sections.

\section{Experimental Feasibility}
\label{sec:feasibility}

The main ingredients required for the proposed control scheme are
available in levitated-particle platforms. Programmable optical tweezers
allow independent trapping and control of multiple nanoparticles
\cite{yan2023ondemand,vijayan2023scalable}, while multi-particle arrays
with site-resolved motion detection have also been demonstrated
\cite{siegel2025arrays,ren2025neuromorphic}. Tunable optical interactions
between spatially separated nanoparticles have been realized
\cite{rieser2022tunable,vijayan2024cavity}, including controllable
nonreciprocal and non-Hermitian dynamics
\cite{reisenbauer2024nonhermitian}. Integrating these capabilities into a
periodically driven nonreciprocal ring remains an experimental challenge,
but the $N=100$ system used in the main calculations is not required to
observe the spatial-phase dynamics.

Our finite-size calculations substantially lower this requirement. A
three-site ring already supports the counterpropagating finite-wave-vector
pair needed for temporal locking, persistent spatial-phase drift, and the
phase--spectrum correspondence, providing a minimal few-particle
demonstrator of the rotating-state dynamics. A four-site ring is the
smallest geometry that additionally retains the commensurate
$q_0=\pi/2$ sector used for the pinned reference state. The same
correspondence persists over the tested finite sizes; the size dependence
and few-site parameter checks are reported in Sec.~S7 of the Supporting
Information.

The relevant observables can be obtained directly from site-resolved
trajectories $x_i(t)$. Using the two-branch decomposition in
Eq.~\eqref{eq:two-branch}, demodulation near half the drive frequency
yields $B_\pm(t)$ and hence the collective phases $\Sigma(t)$ and
$\Delta(t)$. A rotating state is therefore identified by temporal locking
of $\Sigma$ together with a steadily evolving $\Delta$. The same
trajectories yield the laboratory frame spectrum $S(q,\omega)$, and
agreement between the reconstructed phase drift velocity and the spectral
splitting in Eq.~\eqref{eq:spectral-relation} provides an independent
consistency check. Reversing both quantities under
$\nu\rightarrow-\nu$ would provide a direct signature of nonreciprocal
control.

The illustrative dimensional mapping gives experimentally resolvable
frequency scales. For the representative parameters considered here, the
rotating state produces a branch separation of order $1\,\mathrm{kHz}$,
comparable to the mode splittings and spectral resolution already achieved
in coupled levitated-particle experiments
\cite{vijayan2024cavity,reisenbauer2024nonhermitian}. Recent experiments
have also demonstrated tunable nonreciprocal optical interactions and
collective dynamics between levitated particles
\cite{reisenbauer2024nonhermitian,liska2024pt}, as well as independently
controlled and simultaneously imaged multiparticle arrays
\cite{siegel2025arrays}. The detailed dimensional conversion and parameter
estimates are given in the Supporting Information. The main remaining
challenge is therefore to combine these capabilities in a closed
few-particle ring while maintaining sufficiently uniform trapping and
stable relative phases over the observation window.

\section{Conclusion and Outlook}

Nonreciprocity provides a mechanism to control the spatial dynamics of a
finite-wave-vector Floquet state while preserving collective temporal
locking. In the driven levitated-particle ring studied here, asymmetric
coupling polarizes the paired finite-$q$ modes and unlocks the spatial
phase into persistent directional drift. The phase drift velocity is
encoded in the antisymmetric splitting of the dominant spectral branches
and reverses with the coupling asymmetry. The pinned regime retains a
stationary spatial phase and laboratory-frame period-two response, whereas the
rotating regime remains temporally locked but lacks strict period doubling
at a fixed site and therefore forms a distinct spatiotemporally ordered
state. This separation provides an additional degree of control over
spatiotemporal Floquet dynamics, allowing the direction and speed of
spatial-phase drift to be tuned without disrupting collective temporal
synchronization.

The spatial phase specifies the position of the finite-wave-vector
pattern around the ring and can therefore be viewed as a collective phase
coordinate. Its dynamics reflects a competition between the directional
bias induced by nonreciprocity and the nonlinear and lattice-induced
tendency to keep the pattern pinned. Once this pinning is overcome, the
pattern rotates persistently, with a velocity that varies continuously
with the drive and coupling parameters. This behavior is reminiscent of
sliding dynamics in spatially modulated media
\cite{sheng2024terahertz,guillet2025melting}. 
The spatial-phase coordinate here, however, is not a Goldstone mode
associated with the spontaneous breaking of an exact continuous
translation symmetry
\cite{kiselev2025symmetry}. This distinction motivates further questions
about how defects, disorder, boundaries, and spatial commensurability
modify pattern rotation, and whether stronger pinning or modified
interactions can stabilize commensurate spatiotemporal locking \cite{funami2024frequency}.

A particularly important open question is whether nonreciprocity can
generate persistent pattern rotation without destroying exact subharmonic
temporal order. In the rotating state studied here, the rotating pattern
removes the strict period-two response at a fixed site. Realizing a
nonreciprocal discrete time crystal that combines exact temporal
period-doubling with controllable pattern rotation therefore remains an
open problem. Recent studies have shown that non-Hermitian or
nonreciprocal interactions can stabilize time-crystalline behavior in
discrete, continuous, and classical settings
\cite{yousefjani2025nonhermitian,karthiga2025reviving,
li2026fragmentation}. Related progress in Floquet photonics has also
demonstrated controllable rotational dynamics in space-time-modulated
resonator networks
\cite{nasari2026rotational}. Whether such mechanisms can preserve strict
subharmonic order while simultaneously supporting controlled spatial
motion, particularly in quantum many-body systems, is an intriguing
direction for further study.

\begin{acknowledgments}

This work is supported by the National Natural Science Foundation of China (Grant No. 12441502) and the Beijing Institute of Technology Research Fund Program
under Grant No. 2024CX01015.

\end{acknowledgments}

\section*{Data availability}

The data that support the findings of this study are available from the corresponding author upon reasonable request.

\bibliography{references}

\clearpage
\onecolumngrid

\begin{center}
{\Large\bfseries Supporting Information for\\[0.5ex]
``Nonreciprocal Control of Spatiotemporal Floquet Dynamics in Levitated Optomechanical Arrays''\par}
\vspace{1.0em}
{\large Guangyu Zhang$^{1,2}$, Qihao Guo$^{3}$, and Zhangqi Yin$^{1,2}$\par}
\vspace{0.5em}
{\small
$^{1}$Center for Quantum Technology Research and Key Laboratory of Advanced Optoelectronic Quantum Architecture and Measurements (MOE), School of Physics, Beijing Institute of Technology, Beijing 100081, China\\
$^{2}$Beijing Key Laboratory of Quantum Matter State Control and Ultra-Precision Measurement Technology, School of Physics, Beijing Institute of Technology, Beijing 100081, China\\
$^{3}$Department of Applied Physics, Xi'an Jiaotong University, Xi'an, Shaanxi 710049, China\par}
\vspace{0.5em}
{\small October 15, 2026\par}
\end{center}
\vspace{1.5em}

\setcounter{figure}{0}
\renewcommand{\thefigure}{S\arabic{figure}}
\setcounter{table}{0}
\renewcommand{\thetable}{S\arabic{table}}
\setcounter{equation}{0}
\renewcommand{\theequation}{S\arabic{equation}}
\setcounter{section}{0}
\renewcommand{\thesection}{S\arabic{section}}

This Supporting Information gives the model construction (Sec.~\ref{sec:si-model}),
linear Floquet theory (Sec.~\ref{sec:si-linear}), numerical protocol and
observable extraction (Sec.~\ref{sec:si-numerics}), nonlinear two-branch
dynamics (Sec.~\ref{sec:si-nonlinear}), a finite-detuning test of the pinned
period-two response (Sec.~\ref{sec:si-rigidity}), an illustrative
dimensional mapping (Sec.~\ref{sec:si-mapping}), and the few-site size scan
(Sec.~\ref{sec:si-few-site}).

\section{Model construction and nondimensionalization}
\label{sec:si-model}

We consider a periodic ring of identical particles with dimensional
displacements $X_i(t_{\rm phys})$ and mass $M$. A convenient dimensional form
of the driven equation is
\begin{align}
M\frac{\mathrm d^2X_i}{\mathrm dt_{\rm phys}^2}
+M\Gamma\frac{\mathrm dX_i}{\mathrm dt_{\rm phys}}
={}&-\kappa_0\left[1+A_D\cos(\Omega_D^{\rm phys}t_{\rm phys})\right]
X_i e^{-2X_i^2/\ell^2}-\alpha X_i^3 \nonumber\\
&+K_R(X_{i-1}-X_i)+K_L(X_{i+1}-X_i)+\Xi_i(t_{\rm phys}).
\label{eq:si-dimensional-model}
\end{align}
Here $\ell$ is the characteristic displacement scale of the Gaussian trap,
$\kappa_0$ is its small-displacement stiffness, and $\alpha$ denotes an
additional cubic confinement. The directed nearest-neighbor couplings are
written as
\begin{equation}
K_R=K_0(1+\nu),\qquad K_L=K_0(1-\nu),
\label{eq:si-dimensional-couplings}
\end{equation}
so that $\nu=0$ is reciprocal, while changing the sign of $\nu$ interchanges
the left- and right-directed couplings. Periodic boundary conditions identify
$X_{i+N}=X_i$.

The local baths are taken to be independent and Markovian,
\begin{equation}
\left\langle\Xi_i(t_{\rm phys})\Xi_j(t'_{\rm phys})\right\rangle
=2M\Gamma k_{\rm B}T_{\rm phys}\,
\delta_{ij}\delta(t_{\rm phys}-t'_{\rm phys}).
\label{eq:si-dimensional-noise}
\end{equation}
The noise represents the local dissipative environment. The directed
couplings need not obey detailed balance and are not assigned an additional
fluctuation--dissipation relation.

To connect this dimensional description to the variables used in the
simulations, we introduce the dimensionless displacement and time
\begin{equation}
X_i=\ell x_i,\qquad t=\omega_{\rm ref}t_{\rm phys},
\label{eq:si-scaling}
\end{equation}
where $t$ is dimensionless. The associated energy scale is
\begin{equation}
E_0=M\ell^2\omega_{\rm ref}^2.
\label{eq:si-energy-scale}
\end{equation}
Dividing Eq.~\eqref{eq:si-dimensional-model} by
$M\ell\omega_{\rm ref}^2$ gives
\begin{align}
\ddot{x}_i+\gamma\dot{x}_i={}&
-\omega_0^2\left[1+A_D\cos(\Omega_Dt)\right]x_i e^{-2x_i^2}
-\beta x_i^3 \nonumber\\
&+k_R(x_{i-1}-x_i)+k_L(x_{i+1}-x_i)+\xi_i(t),
\label{eq:si-model}
\end{align}
where overdots denote derivatives with respect to dimensionless time and
\begin{align}
\gamma&=\frac{\Gamma}{\omega_{\rm ref}},
&\omega_0^2&=\frac{\kappa_0}{M\omega_{\rm ref}^2},
&\Omega_D&=\frac{\Omega_D^{\rm phys}}{\omega_{\rm ref}},\nonumber\\
\beta&=\frac{\alpha\ell^2}{M\omega_{\rm ref}^2},
&k_{R,L}&=\frac{K_{R,L}}{M\omega_{\rm ref}^2},
&T&=\frac{k_{\rm B}T_{\rm phys}}{E_0}.
\label{eq:si-dimensionless-parameters}
\end{align}
The dimensionless noise therefore satisfies
\begin{equation}
\left\langle\xi_i(t)\xi_j(t')\right\rangle
=2\gamma T\,\delta_{ij}\delta(t-t').
\label{eq:si-noise}
\end{equation}
Throughout the paper, $T$ denotes this dimensionless bath temperature
(noise strength), not a temperature expressed directly in kelvin. We choose
$\omega_{\rm ref}$ such that $\omega_0=1$ in the simulations.

Table~\ref{tab:si-parameters} lists the baseline dimensionless parameters.
The drive amplitude $A_D$, nonreciprocity $\nu$, temperature $T$, and system
size $N$ are varied in the protocols specified in Supporting Information
Sec.~\ref{sec:si-numerics}.

\begin{center}
\refstepcounter{table}
\label{tab:si-parameters}
\noindent{\small TABLE~\thetable. \textbf{Dimensionless definitions and baseline parameters.}\par}
\vspace{0.6ex}
\begin{ruledtabular}
\begin{tabular}{lcl}
Quantity & Baseline value & Definition\\
\hline
$N$ & $100$ & Sites on the periodic ring\\
$\omega_0$ & $1$ & Onsite frequency and frequency unit\\
$\gamma$ & $0.15$ & Viscous damping rate\\
$\beta$ & $1$ & Cubic onsite coefficient\\
$k_0$ & $0.1$ & Mean coupling, $k_{R,L}=k_0(1\pm\nu)$\\
$\Omega_D$ & $2.07$ & Drive angular frequency\\
$\delta_p$ & $0.07$ & Drive detuning, $\Omega_D-2\omega_0$\\
$T$ & $10^{-3}$ & Baseline dimensionless bath temperature\\
\end{tabular}
\end{ruledtabular}
\end{center}

Here $\delta_p$ denotes the full-drive detuning
$\Omega_D-2\omega_0$; the corresponding half-drive detuning is
$\delta_p/2$.

\section{Linear Floquet analysis and paired rotating-wave approximation}
\label{sec:si-linear}

Linearizing Eq.~\eqref{eq:si-model} around $x_i=0$ and using
\begin{equation}
x_i(t)=\frac{1}{\sqrt N}\sum_q x_q(t)e^{iqi},
\qquad q=\frac{2\pi m}{N},
\label{eq:si-spatial-fourier}
\end{equation}
gives the directed coupling contribution
\begin{equation}
k_R(e^{-iq}-1)+k_L(e^{iq}-1)
=2k_0(\cos q-1)-2ik_0\nu\sin q.
\label{eq:si-coupling}
\end{equation}
Each Fourier component obeys
\begin{equation}
\ddot{x}_q+\gamma\dot{x}_q+
\left[\Omega_q^2+i\chi_q+A_D\omega_0^2\cos(\Omega_Dt)\right]x_q=0,
\label{eq:si-floquet-mode}
\end{equation}
with
\begin{equation}
\Omega_q^2=\omega_0^2+2k_0(1-\cos q),
\qquad
\chi_q=2k_0\nu\sin q.
\label{eq:si-linear-parameters}
\end{equation}
The reciprocal contribution is even in $q$, whereas the nonreciprocal term
is odd.

We first determine the instability without a rotating-wave approximation.
Writing $\bm z_q=(x_q,\dot x_q)^{\mathsf T}$ gives
\begin{equation}
\dot{\bm z}_q=\mathsf L_q(t)\bm z_q,
\qquad
\mathsf L_q(t)=
\begin{pmatrix}
0&1\\
-\Omega_q^2-i\chi_q-A_D\omega_0^2\cos(\Omega_Dt)&-\gamma
\end{pmatrix}.
\label{eq:si-first-order}
\end{equation}
The one-period Floquet propagator is
\begin{equation}
\mathsf F_q=\mathcal T
\exp\!\left[\int_0^{T_D}\mathsf L_q(t)\,\mathrm dt\right],
\qquad
\mathsf F_q\bm u_{\alpha,q}=\mu_\alpha(q)\bm u_{\alpha,q}.
\label{eq:si-floquet-propagator}
\end{equation}
The maximal Floquet growth exponent is
\begin{equation}
G(q)=\max_\alpha\frac{\ln|\mu_\alpha(q)|}{T_D}.
\label{eq:si-floquet-growth}
\end{equation}
Because $\mathsf L_{-q}(t)=\mathsf L_q^*(t)$, one has
$\mathsf F_{-q}=\mathsf F_q^*$ and hence $G(-q)=G(q)$. Thus the exact
linear instability remains inversion symmetric in its growth rate even
though the paired eigenvector composition is nonreciprocally polarized.

To interpret the exact instability analytically, near the principal
parametric resonance we write
\begin{equation}
x_i(t)=\psi_i(t)e^{i\omega t}+\psi_i^*(t)e^{-i\omega t},
\qquad \omega=\frac{\Omega_D}{2},
\label{eq:si-rwa-ansatz}
\end{equation}
with $|\dot\psi_i|\ll\omega|\psi_i|$. Parametric driving couples
$\psi_q$ to $\psi_{-q}^*$, giving
\begin{equation}
\frac{\mathrm d}{\mathrm dt}
\begin{pmatrix}\psi_q\\\psi_{-q}^*\end{pmatrix}
=\mathsf M_q
\begin{pmatrix}\psi_q\\\psi_{-q}^*\end{pmatrix},
\qquad
\mathsf M_q=
\begin{pmatrix}
-\gamma/2-g_q+i\delta_q&p\\
p&-\gamma/2+g_q-i\delta_q
\end{pmatrix},
\label{eq:si-paired-matrix}
\end{equation}
where
\begin{equation}
p=\frac{A_D\omega_0^2}{4\omega},\qquad
g_q=\frac{k_0\nu\sin q}{\omega},\qquad
\delta_q=\frac{\omega_0^2-\omega^2+2k_0(1-\cos q)}{2\omega}.
\label{eq:si-rwa-coefficients}
\end{equation}
The paired-RWA eigenvalues are
\begin{equation}
\lambda_\pm(q)=-\frac{\gamma}{2}
\pm\sqrt{p^2+(-g_q+i\delta_q)^2}.
\label{eq:si-eigenvalues}
\end{equation}
At onset, $\operatorname{Re}\lambda_+=0$, which gives
\begin{align}
p_{\rm on}^2(q,\nu)&=\delta_q^2+\frac{\gamma^2}{4}
-g_q^2\left(1+\frac{4\delta_q^2}{\gamma^2}\right),\nonumber\\
A_D^{\rm on}(q,\nu)&=\frac{4\omega}{\omega_0^2}
\sqrt{\max[p_{\rm on}^2(q,\nu),0]}.
\label{eq:si-onset}
\end{align}
The minimizing allowed wave vector gives the paired-RWA estimate of the
finite-$q$ onset. The nonlinear simulations do not impose this wave vector;
$q_0$ is remeasured after saturation.

The same paired matrix also exposes the internal directional structure of
the unstable mode. Let the dominant right eigenvector of $\mathsf M_q$ be
\begin{equation}
\bm u_+(q)=(u_q,v_q)^{\mathsf T},
\qquad \mathsf M_q\bm u_+=\lambda_+\bm u_+.
\label{eq:si-linear-eigenvector}
\end{equation}
Its branch ratio and normalized polarization are
\begin{equation}
r_q\equiv\frac{v_q}{u_q}
=\frac{\lambda_++\gamma/2+g_q-i\delta_q}{p},
\qquad
m_{\rm lin}(q)=
\frac{|u_q|^2-|v_q|^2}{|u_q|^2+|v_q|^2}
=\frac{1-|r_q|^2}{1+|r_q|^2}.
\label{eq:si-linear-polarization}
\end{equation}
The quantity $m_{\rm lin}$ is independent of eigenvector normalization.
It changes sign when the favored member of the $(+q,-q)$ pair is exchanged.
At $q=\pi$, the lattice modes $+q$ and $-q$ are identical modulo a reciprocal
lattice vector and $g_\pi=0$; the two-branch polarization is therefore
degenerate and should not be interpreted as an independent directional pair.

\begin{figure}[t]
\centering
\includegraphics[width=0.90\textwidth]{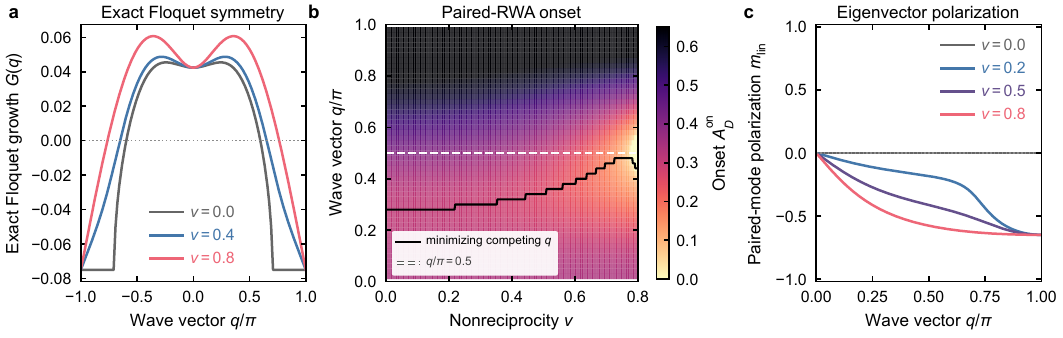}
\caption{\textbf{Exact Floquet instability and paired-mode response.}
\textbf{a}, Maximal growth exponent $G(q)$ obtained from the full
time-periodic linear problem. \textbf{b}, Paired-RWA onset
$A_D^{\rm on}(q,\nu)$ on the allowed positive-$q$ grid for $N=100$.
\textbf{c}, Polarization $m_{\rm lin}$ of the dominant paired eigenvector.
The exact Floquet calculation determines the unstable finite-$q$ sector,
whereas the paired-RWA exposes the nonreciprocal branch polarization.}
\label{fig:si-linear}
\end{figure}
\FloatBarrier

\section{Numerical protocol and phase extraction}
\label{sec:si-numerics}

We write $v_i=\dot x_i$ and denote the deterministic force excluding
damping by $F_i(\bm x,t)$. During one integration step the force is evaluated
at the beginning of the step, while the linear damping and Gaussian noise are
integrated analytically:
\begin{align}
v_i^{m+1}={}&e^{-\gamma\Delta t}v_i^m
+\frac{1-e^{-\gamma\Delta t}}{\gamma}F_i(\bm x^m,t_m)
+\sqrt{T(1-e^{-2\gamma\Delta t})}\,\zeta_i^m,\nonumber\\
x_i^{m+1}={}&x_i^m+v_i^{m+1}\Delta t.
\label{eq:si-integrator}
\end{align}
The $\zeta_i^m$ are independent standard normal variables. At $T=0$ the
stochastic term is omitted. This is the production exponential
damping/noise update with a frozen deterministic force, followed by the
position drift; it is not a BAOAB or Runge--Kutta integrator. Unless stated
otherwise, $\Delta t=T_D/40$. Initial positions and velocities are drawn
independently from zero-mean normal distributions with standard deviation
$0.1$. Separate random seeds control the initial condition and Langevin
noise when the protocol requires them to be varied independently.

Table~\ref{tab:si-protocols} records the simulations entering the main-text
figures. Preparation cycles are discarded before recording, whereas the
analysis column states the portion of each recorded trajectory entering the
reported observable. The theoretical results in Figure~2 do not require
nonlinear trajectory integration.

\begin{center}
\refstepcounter{table}
\label{tab:si-protocols}
\noindent{\small TABLE~\thetable. \textbf{Main-text numerical protocols.}
The baseline parameters not listed explicitly are those of
Table~\ref{tab:si-parameters}.\par}
\vspace{0.6ex}
\scriptsize
\resizebox{\textwidth}{!}{%
\begin{tabular}{llllllll}
\toprule
Figure & Scan or representative parameters & $T$ & Realizations & Preparation & Recorded & Analysis & Steps/cycle\\
\midrule
3a & $A_D=0.25{:}0.01{:}0.65$, $\nu=0{:}0.02105{:}0.80,0.825,0.850$ & $10^{-3}$ & 5/grid point & 2100 & 900 & nested 300/600/900 & 40\\
3b,c pinned & $(A_D,\nu)=(0.54,0.2947)$ & $10^{-3}$ & 1 & 2000 & 1000 & full record & 40\\
3b,c rotating & $(A_D,\nu)=(0.49,0.80)$ & $10^{-3}$ & 1 & 1900 & 1100 & full record & 40\\
4a & same pinned/rotating representatives & 0 & 1 each & 2000 & 8000 & full record & 80\\
4b & $11\times9$ predefined $(A_D,\nu)$ grid & $10^{-3}$ & 1/candidate & 1200 & 1200 & full record & 40\\
5a & $A_D=0.40,0.50,0.60$, $\nu=-1{:}0.05{:}1$ & $10^{-3}$ & 5/grid point & 2100 & 900 & nested 300/600/900 & 40\\
5b & $\nu:+0.8\to-0.8\to+0.8$, $A_D=0.49$ & $10^{-3}$ & 1 & 1800 & 1350 & three settled windows & 40\\
6a & 200-cycle pulses in $A_D$, $\nu$, or $T$ & baseline/pulse & 20/protocol & 1800 & 1600 & pre/pulse/recovery & 40\\
6b & $N=50,100,150,200$ at $(A_D,\nu)=(0.49,0.80)$ & $10^{-3}$ & 5/size & 1200 & 1200 & full record & 40\\
\bottomrule
\end{tabular}
}

\vspace{1.2ex}
\begin{minipage}{0.98\textwidth}
\textbf{Analysis and selection.}
Unless specified otherwise, the production step is $\Delta t=T_D/40$, and
the half-drive-frequency envelope uses a normalized Hann window of width
$2T_D$.
Figure~3a uses nested 300-, 600-, and 900-cycle spatial-phase fits and displays the
median normalized spatial-phase drift velocity $v_\Delta$ only when at least three of five seeds are lock eligible.
For Figure~3b,c, the stripe plots contain ten snapshots separated by five cycles
in a 30-site window. Figure~4a uses a Hann time window, physical FFT bins, and
no zero padding. Of the 99 predefined Figure~4b candidates, 48 pass the
$r_\Sigma\geq0.98$, branch-quality, and peak-resolution criteria; fourfold
padding is used only for peak localization. Figure~5a uses the same nested
spatial-phase-fit windows as Figure~3a. The three Figure~5b fit windows are protocol
cycles 80--320, 430--800, and 930--1300. Figure~6a reports the local velocity
relative to each realization's pre-pulse value. Figure~6b selects $q_0$
independently for every size and seed and evaluates both residuals from the
final 1200 cycles.
\end{minipage}
\end{center}

From each retained trajectory, we first extract the complex
half-drive-frequency envelope
\begin{equation}
\psi_i(t)=\operatorname{LPF}
\left[x_i(t)e^{-i\Omega_Dt/2}\right],
\label{eq:si-envelope}
\end{equation}
where LPF denotes a normalized Hann convolution of width $2T_D$. From the
late-time stroboscopic envelope we form
\begin{equation}
S(q)=\left\langle
\left|\frac{1}{\sqrt N}\sum_i\psi_i(n)e^{-iqi}\right|^2
\right\rangle_n.
\label{eq:si-spatial-spectrum}
\end{equation}
The positive carrier $q_0$ maximizes the paired power $S(q)+S(-q)$ over the
allowed nonzero modes. The carrier is re-evaluated for every trajectory and
is not imposed from linear theory.

The selected carrier is then removed to obtain local branch fields. Applying
a periodic Gaussian spatial filter gives
\begin{equation}
B_{\pm,i}(n)=\sum_{r=-6}^{6}K_r\psi_{i+r}(n)
e^{\mp iq_0(i+r)},
\qquad
K_r\propto e^{-r^2/(2\sigma_x^2)},\quad \sigma_x=2,
\label{eq:si-branch-filter}
\end{equation}
with $\sum_rK_r=1$. We define
\begin{equation}
\Psi_{\Sigma,i}=(B_{+,i}B_{-,i})^*,
\qquad
\Psi_{\Delta,i}=(B_{+,i}B_{-,i}^*)^*.
\label{eq:si-phase-fields}
\end{equation}
We fix the overlap reference directly as
\begin{equation}
O_{\rm max}=\max_{j,n}|B_{+,j}(n)B_{-,j}(n)|.
\label{eq:si-overlap-maximum}
\end{equation}
At each time, the valid-site set is
\begin{equation}
\mathcal V(n)=\left\{i:
|B_{+,i}(n)B_{-,i}(n)|>\epsilon_{\rm ov}\,O_{\rm max}\right\},
\qquad \epsilon_{\rm ov}=10^{-3},
\label{eq:si-valid-sites}
\end{equation}
The overlap-weighted global phases are then
\begin{align}
Z_\Sigma(n)&=\sum_{i\in\mathcal V(n)}\Psi_{\Sigma,i}(n),
&\Sigma_g(n)&=\operatorname{unwrap}\arg Z_\Sigma(n),\nonumber\\
Z_\Delta(n)&=\sum_{i\in\mathcal V(n)}\Psi_{\Delta,i}(n),
&\Delta_g(n)&=\operatorname{unwrap}\arg Z_\Delta(n).
\label{eq:si-global-phases}
\end{align}

These branch fields define the temporal-locking diagnostic. The instantaneous
overlap-weighted circular coherence is
\begin{equation}
r_\Sigma(n)=
\frac{|Z_\Sigma(n)|}
{\sum_{i\in\mathcal V(n)}|B_{+,i}(n)B_{-,i}(n)|},
\qquad
r_\Sigma=\langle r_\Sigma(n)\rangle_{\rm late}.
\label{eq:si-coherence}
\end{equation}
A trajectory in the main phase-map scan is classified as temporally locked
when
\begin{equation}
r_\Sigma\geq0.97,
\qquad
\left|\frac{\mathrm d\Sigma_g}{\mathrm dn}\right|
\leq5\times10^{-3}\ \mathrm{rad/cycle}.
\label{eq:si-locking-criterion}
\end{equation}
The first requirement tests spatial coherence of the local temporal phases;
the second excludes systematic global temporal-phase winding. For a narrow
circular distribution, $r_\Sigma=0.97$ corresponds to
\begin{equation}
\sigma_{\Sigma,{\rm circ}}=\sqrt{-2\ln r_\Sigma}
\simeq0.247\ \mathrm{rad}\simeq14.1^\circ.
\label{eq:si-circular-spread}
\end{equation}
The retained window in the phase-map scan is 900 cycles, for which one full
$2\pi$ phase winding corresponds to $2\pi/900=6.98\times10^{-3}$ rad/cycle.
The adopted drift tolerance is below this value. These numbers are fixed
operational tolerances, not thermodynamic critical values.

At each $(A_D,\nu)$ point, at least three of five trajectories must satisfy
the locking and data-quality conditions before a phase drift velocity is shown
in Figure~3a. The data-quality conditions require a resolved finite-$q$ band,
a usable mixed branch, and a sufficient valid-site fraction. Grid points
without the required seed consensus are hatched. The stricter condition
$r_\Sigma\geq0.98$ used for the spectral comparison in Figure~4b is an
additional coherence selection for clean peak extraction; it does not define
a second temporal-locking boundary.

Spatial-phase drift and its independent spectral counterpart are quantified next.
The pattern displacement in lattice sites is
\begin{equation}
X(n)-X(n_0)=\frac{\Delta_g(n)-\Delta_g(n_0)}{2q_0}.
\label{eq:si-registry-sites}
\end{equation}
The spatial-phase drift rate $\mathrm d\Delta_g/\mathrm dn$ defines the
normalized spatial-phase drift velocity, measured in pattern wavelengths
per drive cycle:
\begin{equation}
\boxed{
v_\Delta=\frac{1}{4\pi}\frac{\mathrm d\Delta_g}{\mathrm dn}}
\qquad
\text{(pattern wavelengths per drive cycle)}.
\label{eq:si-registry-velocity}
\end{equation}
Hereafter $v_\Delta$ is referred to simply as the phase drift velocity.
The corresponding real-space rate is a mode-dependent conversion,
\begin{equation}
\frac{\mathrm dX}{\mathrm dn}=\frac{2\pi}{q_0}v_\Delta.
\label{eq:si-real-space-conversion}
\end{equation}

The joint spectrum is computed from the complete laboratory-frame record,
after subtracting the sitewise mean and applying a Hann window in time:
\begin{equation}
S(q,\omega)=\left|
\sum_{i,n}x_i(t_n)e^{-i(qi-\omega t_n)}
\right|^2.
\label{eq:si-joint-spectrum}
\end{equation}
The representative spectra in Figure~4a use 8000 retained cycles, 80 samples
per cycle, physical Fourier bins, and no zero padding. For a temporally locked
rotating state,
\begin{equation}
\frac{\omega_++\omega_-}{2\Omega_D}=\frac{1}{2},
\qquad
\delta\omega_{\rm spec}
\equiv\frac{|\omega_+-\omega_-|}{2\Omega_D}=|v_\Delta|.
\label{eq:si-spectral-relation}
\end{equation}

For Figure~4b we simulate the predefined grid
\begin{equation}
A_D\in\{0.30,0.33,\ldots,0.60\},\qquad
\nu\in\{0.40,0.45,\ldots,0.80\}.
\label{eq:si-relation-grid}
\end{equation}
The 99 candidates are filtered before comparison with the predicted relation.
The accepted 48 states satisfy $r_\Sigma\geq0.98$, the drift condition in
Eq.~\eqref{eq:si-locking-criterion}, $|v_\Delta|>10^{-3}$ wavelengths/cycle,
the finite-$q$ and branch-quality requirements, and independent resolution
of the two spectral peaks. The horizontal uncertainty is propagated from the
standard error of the fitted spatial-phase slope. The vertical uncertainty
is one half of an independent Fourier bin.

For the finite-size test, the splitting and center-locking residuals are
\begin{equation}
\epsilon_{\rm split}(N)=
\frac{\delta\omega_{\rm spec}(N)-|v_\Delta(N)|}
{\delta\omega_{\rm spec}(N)},
\qquad
\epsilon_c(N)=\frac{\omega_+(N)+\omega_-(N)}{2\Omega_D}-\frac{1}{2}.
\label{eq:si-residuals}
\end{equation}

Finally, the control and robustness tests use the following protocols. For
Figure~5b, the rotating state is first prepared for 1800 cycles at
$A_D=0.49$, $\nu=0.80$, and $T=10^{-3}$. The subsequent protocol applies
$+\nu_0$ for 350 cycles, $-\nu_0$ for 500 cycles, and $+\nu_0$ for 500 cycles,
where $\nu_0=0.80$. Fits exclude the switching transients and use cycles
$80$--$320$, $430$--$800$, and $930$--$1300$ of the recorded protocol.

For Figure~6a, each realization is first equilibrated for 1800 cycles. The
recorded sequence contains 400 pre-pulse cycles, a 200-cycle perturbation,
and 1000 recovery cycles. The three perturbations are
$A_D:0.49\to0.60$, $\nu:0.80\to0.50$, and
$T:10^{-3}\to0.04$. Twenty independent pairs of initial-condition and
noise seeds are used for each protocol. Local velocities are converted to
wavelengths per cycle and referenced to the pre-pulse value of the same
trajectory before calculating the median and interquartile range.

For Figure~6b, $N=50,100,150,200$ are simulated at
$(A_D,\nu,T)=(0.49,0.80,10^{-3})$ with five seeds per size. Each trajectory
contains 2400 cycles, of which the final 1200 are retained. The selected
mode is always an allowed ring momentum $q_m=2\pi m/N$; changes in $q_0$ with
$N$ therefore reflect finite-size momentum quantization rather than a loss of
the splitting relation.
\FloatBarrier

\subsection{Numerical validation}
\label{sec:si-validation}

The production time step is $T_D/40$. Repeating the representative rotating
trajectory with 20, 40, and 80 updates per drive cycle leaves the fitted
wavelength-normalized phase drift velocity unchanged within the seed and
window-to-window variation. The same conclusion is obtained when the start
of the retained fitting window is shifted while its late endpoint is held
fixed.

To test the robustness of the classification itself, the phase-map mask was
recomputed from the cached seed-resolved
metrics without rerunning the dynamics. On the original $0\leq\nu\leq0.80$
grid, the baseline conditions populate 1432 of 1599 points. The two additional
columns at $\nu=0.825$ and $0.850$ add 68 accepted points, giving 1500 of
1681 for Figure~3a. On the original grid, changing the coherence threshold from 0.97 to
0.965 or 0.975 gives mask agreements of $98.9\%$ and $96.3\%$, respectively,
with the changes concentrated near the edge of the accepted region. Varying
the temporal-drift cutoff from $10^{-3}$ to $2\times10^{-2}$ rad/cycle does
not change the displayed map. In fact, among the numerically valid seeds with
$r_\Sigma\geq0.97$, the largest measured temporal drift in the extended scan is
$3.4\times10^{-4}$ rad/cycle. The temporal-drift requirement is therefore a
safety condition against phase winding rather than an active tuning parameter
for Figure~3a.

The finite-size protocol next separates mode quantization from the dynamical
relation. For each $N$, the carrier is selected from $q_m=2\pi m/N$, and the
two residuals in Eq.~\eqref{eq:si-residuals} are evaluated independently for
five seeds. The data in Figure~6b remain close to zero over
$N=50,100,150,200$. Thus changing $N$ changes the available carrier grid but
does not resolve a systematic failure of either the splitting relation or
the locked temporal center. This is a finite-size consistency test over the
accessible sizes, not a proof of thermodynamic long-range order.

Finally, we tested whether the spatial-phase drift in the rotating state locks to a finite-width rational
velocity plateau. At $N=100$, $T=0$, and the baseline model parameters, four
representative $(A_D,\nu)$ cuts were scanned over
$\delta_p\in[-0.50,0.50]$ with spacing $0.01$. Candidate neighborhoods were
then refined over $\delta_p\pm0.02$ with spacing $0.002$ and five initial
conditions per point. A finite-width plateau was required to preserve the
same $q_0$ and low-order rational signature over at least three adjacent grid
points, with agreement of at least three of five initial conditions in the
refined scan. No candidate met this criterion. Additional scans over
$k_0\in[0.04,0.18]$ and $\gamma\in[0.01,0.30]$ likewise produced isolated
near-rational crossings rather than a resolved finite-width plateau.
The stricter coherence selection $r_\Sigma\geq0.98$ was used in these
auxiliary searches. These results support continuous spatial-phase drift over
the scanned cuts and resolution, but do not exclude commensurate locking in
other parameter regions.

\subsection{Statistical Analysis}

The numbers of independent trajectories and the analysis windows for
each main-text figure are listed in Table~\ref{tab:si-protocols}.
Figures~3a and 5a show medians over eligible trajectories from five
seeds per grid point, with interquartile ranges where indicated.
Figure~6a uses 20 paired seed realizations per perturbation and reports
the median and interquartile range; Figure~6b uses five seeds per size
and reports means and standard deviations. Figure~\ref{fig:si-few-site}
shows five seeds per size and parameter point, with medians and interquartile
ranges. The Figure~4b comparison
contains 48 accepted trajectories out of 99 predefined candidate
parameter points; its horizontal error bars are standard errors of the
fitted phase slopes and its vertical error bars represent half a Fourier
bin. These numerical diagnostics do not use null-hypothesis tests or
$p$-values. The fixed selection criteria and treatment of excluded
trajectories are specified above; no additional statistical outlier
removal is applied. Phase extraction and summary statistics are computed
with Python and NumPy.

\section{Nonlinear two-branch dynamics and pinned-state stability}
\label{sec:si-nonlinear}

For the full Gaussian onsite force, the first-harmonic projection can be
written in terms of $z_i=4|\psi_i|^2$ as
\begin{align}
G_i^{(0)}&=f_0(\psi_i)\psi_i,\nonumber\\
G_i^{(D)}&=\frac{1}{2}\left[f_0(\psi_i)\psi_i^*
+f_3(\psi_i)\psi_i^3\right],
\label{eq:si-gaussian-projections}
\end{align}
where
\begin{equation}
f_0=e^{-z_i}[I_0(z_i)-I_1(z_i)],\qquad
f_3=\frac{e^{-z_i}[I_2(z_i)-I_1(z_i)]}{|\psi_i|^2},
\label{eq:si-gaussian-coefficients}
\end{equation}
where $I_m$ is the modified Bessel function and $f_3(0)=-2$. The projected slow flow is
\begin{align}
\dot\psi_i={}&-\frac{\gamma}{2}\psi_i+
\frac{1}{2i\omega}\Big[
\omega^2\psi_i-\omega_0^2G_i^{(0)}-A_D\omega_0^2G_i^{(D)}
-3\beta|\psi_i|^2\psi_i\nonumber\\
&\hspace{32mm}+k_R(\psi_{i-1}-\psi_i)
+k_L(\psi_{i+1}-\psi_i)\Big].
\label{eq:si-full-flow}
\end{align}
This equation retains the full amplitude dependence of the Gaussian force
within the first temporal harmonic.

We next project this slow flow onto the counterpropagating pair selected by
the microscopic dynamics and decompose
\begin{equation}
\psi_i=B_+e^{iq_0i}+B_-e^{-iq_0i}+\psi_i^\perp,
\qquad B_\pm=R_\pm e^{-i\phi_\pm}.
\label{eq:si-branch-expansion}
\end{equation}
Keeping the uniform cubic terms gives
\begin{align}
\dot B_+={}&\left(-\frac{\gamma}{2}-g_{q_0}+i\delta_{q_0}\right)B_+
+pB_-^*
-iu\left[(|B_+|^2+2|B_-|^2)B_+
+\eta_cB_-^2B_+^*\right]+\mathcal R_+,\nonumber\\
\dot B_-={}&\left(-\frac{\gamma}{2}+g_{q_0}+i\delta_{q_0}\right)B_-
+pB_+^*
-iu\left[(|B_-|^2+2|B_+|^2)B_-
+\eta_cB_+^2B_-^*\right]+\mathcal R_-,
\label{eq:si-branch-equations}
\end{align}
where
\begin{equation}
u=\frac{3(2\omega_0^2-\beta)}{2\omega},
\qquad
\eta_c(q_0)=\frac{1}{N}\sum_{j=0}^{N-1}e^{i4q_0j}.
\label{eq:si-nonlinear-coefficients}
\end{equation}
For a perfectly uniform projection, $\eta_c=1$ when
$4q_0=2\pi\ell$ and vanishes otherwise. The residuals $\mathcal R_\pm$
contain spatial gradients, modes outside the selected pair, nonresonant
harmonics, and noise. They are not fitted to the measured phase drift velocity.

To expose the collective coordinates of this pair, we define
\begin{equation}
S=R_+^2+R_-^2,\qquad
m=\frac{R_+^2-R_-^2}{S},\qquad
\Sigma=\phi_++\phi_-,\qquad
\Delta=\phi_+-\phi_-.
\label{eq:si-four-vars}
\end{equation}
Here $S$ is the paired intensity, $m$ is the nonlinear branch imbalance,
$\Sigma$ is the collective temporal phase, and $\Delta$ is the collective
phase difference that determines the spatial phase $\Delta/2$ of the
finite-$q$ pattern. Their geometry follows from
\begin{equation}
B_+B_-=R_+R_-e^{-i\Sigma},\qquad
\frac{B_+}{B_-}=\frac{R_+}{R_-}e^{-i\Delta}.
\label{eq:si-phase-geometry}
\end{equation}
For the real-$p$ cubic closure, the four variables satisfy
\begin{align}
\frac{\dot S}{S}&=-\gamma-2g_{q_0}m
+2p\sqrt{1-m^2}\cos\Sigma+P_S,\nonumber\\
\frac{\dot m}{1-m^2}&=-2g_{q_0}
-\frac{2pm\cos\Sigma}{\sqrt{1-m^2}}
+\eta_cuS\sin(2\Delta)+P_m,\nonumber\\
\dot\Sigma&=-2\delta_{q_0}+3uS
-\frac{2p\sin\Sigma}{\sqrt{1-m^2}}
+\eta_cuS\cos(2\Delta)+P_\Sigma,\nonumber\\
\dot\Delta&=-umS
+\frac{2pm\sin\Sigma}{\sqrt{1-m^2}}
-\eta_cumS\cos(2\Delta)+P_\Delta.
\label{eq:si-four-flow}
\end{align}
The $P$ terms are induced by $\mathcal R_\pm$. Equation~\eqref{eq:si-four-flow}
establishes the mechanism used in the main text: the odd nonreciprocal term
$g_{q_0}$ directly biases the branch imbalance $m$, and $m$ in turn enters
$\dot\Delta$. The parametric term supplies a restoring contribution to
$\Sigma$, so temporal locking does not require a stationary spatial phase.

For commensurate carriers, the $\eta_c$ terms provide a periodic dependence
on the spatial-phase coordinate. Additional finite-ring and multimode corrections
are contained in $P_m$ and $P_\Delta$. These terms can be interpreted as an
effective spatial-phase-restoring landscape competing with the directional bias
transmitted through $m$. We use this only as a depinning interpretation. No
stand-alone Adler equation is fitted to the microscopic trajectories, and no
effective bias or pinning coefficient is used to generate the main-text data.

To determine the actual pinned-state stability boundary quantitatively, we
return to the full Gaussian slow flow. The paired equations describe the
phase structure but do not by themselves locate the stability boundary used
in Figure~3a of the main text. We determine
that boundary from the full Gaussian slow flow, Eq.~\eqref{eq:si-full-flow}.
For the commensurate $q_0=\pi/2$ solution, four complex amplitudes span one
unit cell. We solve the fixed-point equations as eight real equations and
test stability against every Bloch perturbation allowed by the $N=100$ ring.
The pinned solution loses stability when the largest real part of the Bloch
spectrum crosses zero. Pseudo-arclength continuation is used to follow the
boundary through folds.

\begin{figure}[t]
\centering
\includegraphics[width=0.78\textwidth]{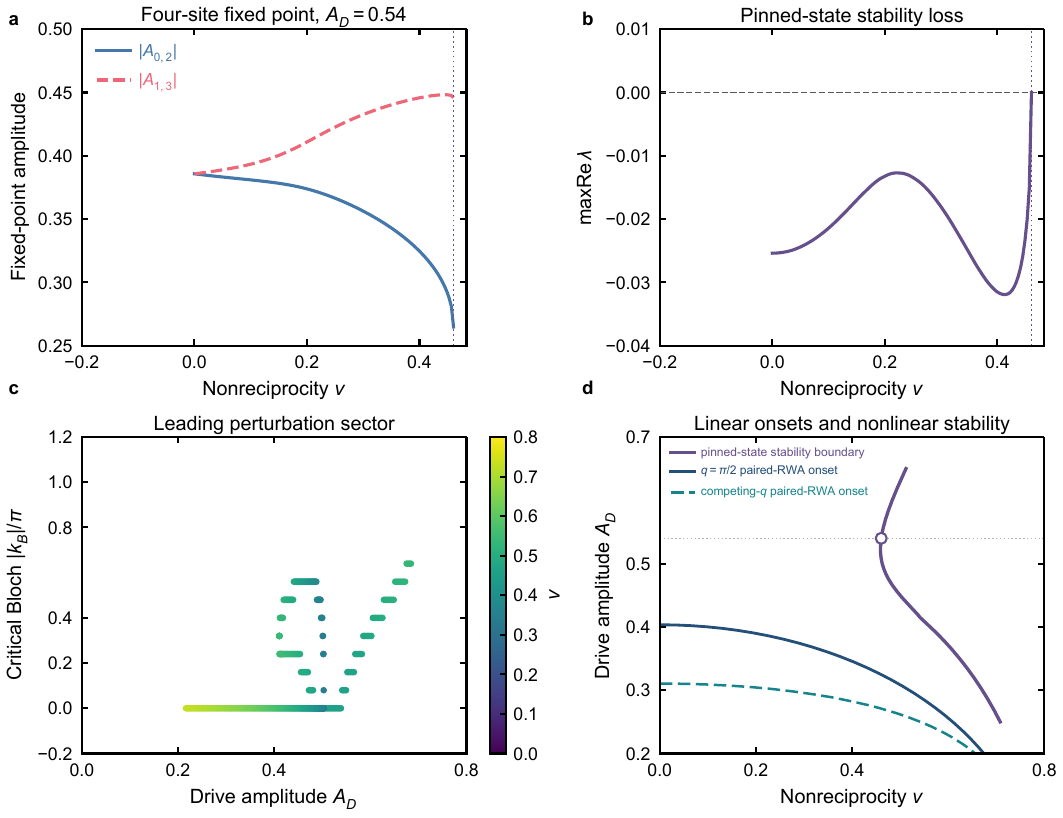}
\caption{\textbf{Nonlinear stability of the commensurate pinned state.}
\textbf{a}, Fixed-point amplitudes of the two symmetry-related site pairs at
$A_D=0.54$. \textbf{b}, Largest real part of the Bloch stability spectrum
along the same cut. \textbf{c}, Wave vector of the leading perturbation.
\textbf{d}, Full-Gaussian pinned-state stability boundary together with the
paired-RWA linear onsets. The full-Gaussian boundary is the line overlaid on
the phase-drift velocity map in Figure~3a of the main text.}
\label{fig:si-pinning}
\end{figure}

\FloatBarrier

\section{Rigidity of the pinned time-crystalline state}
\label{sec:si-rigidity}

We test whether the pinned laboratory-frame period-two response is
confined to one drive frequency. At the Figure~3 pinned representative
coupling $(A_D,\nu)=(0.54,0.2947)$ and $T=10^{-3}$, we scan the full-drive
detuning $\delta_p=\Omega_D-2\omega_0$ with five independent seeds per
sampled value. Each $N=100$ trajectory lasts 2400 drive cycles, and
observables are extracted from the final 900 cycles with 40 integration
steps per cycle. The same branch and phase extraction as in
Sec.~\ref{sec:si-numerics} is used throughout.

We require the established temporal-locking criteria
$r_\Sigma\geq0.97$ and
$|\mathrm d\Sigma_g/\mathrm dn|\leq5\times10^{-3}$ rad/cycle.
For this finite-window pinned test we also require
$|v_\Delta|<10^{-3}$ pattern wavelengths per cycle. The microscopic
period-two diagnostic is the median over the retained window of
$\|\bm x((n+1)T_D)+\bm x(nT_D)\|/\|\bm x(nT_D)\|$.
\FloatBarrier

\begin{figure}[t]
\centering
\includegraphics[width=0.93\textwidth]{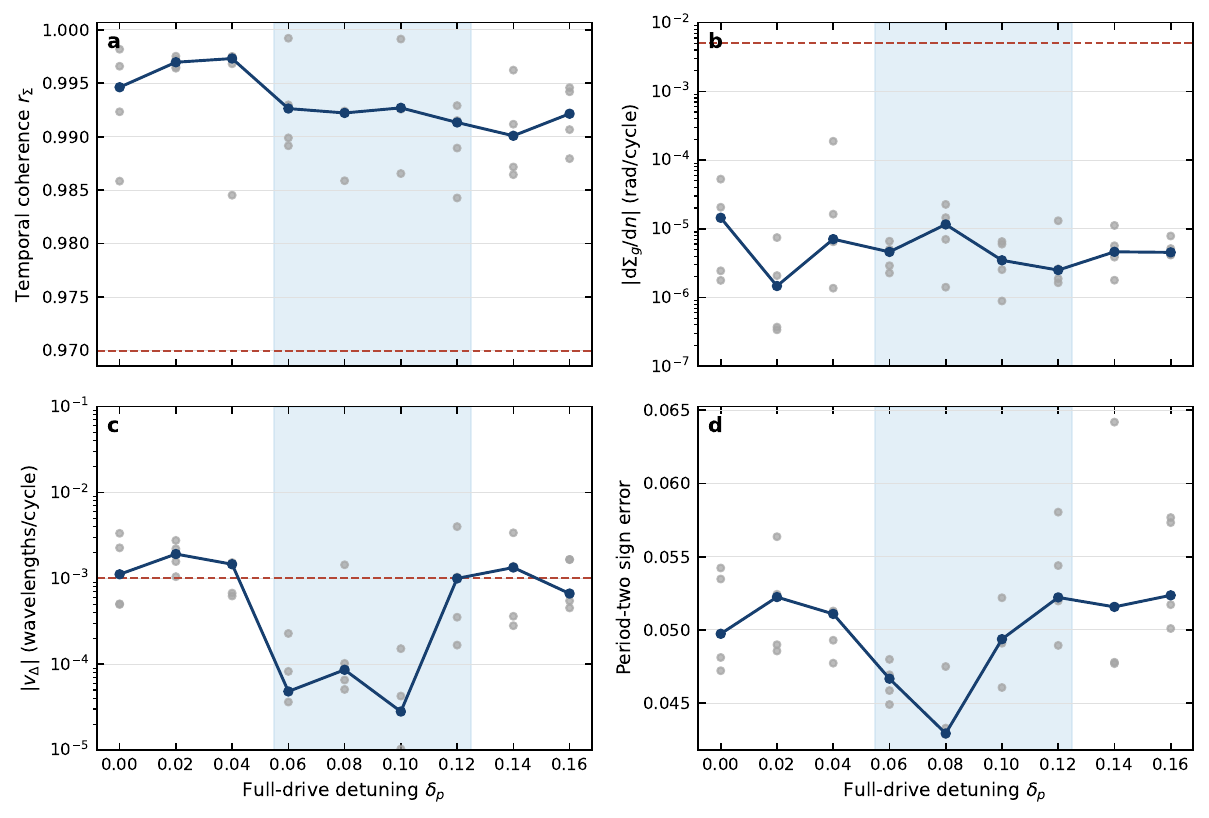}
\caption{\textbf{Sampled finite-detuning rigidity of the pinned response.}
Five seeds at each full-drive detuning $\delta_p$ are shown as light points;
dark markers give medians. The shaded region covers the four adjacent
sampled points $\delta_p=0.06,0.08,0.10,0.12$ for which at least three of
five seeds simultaneously meet the temporal-locking and pinned spatial-phase
criteria. Panels show temporal coherence, global temporal-phase drift,
absolute phase drift velocity, and laboratory-frame consecutive-cycle sign
error. Dashed lines mark the applicable operational thresholds. The
shading is not an estimate of the locking-region boundary.}
\label{fig:si-detuning-rigidity}
\end{figure}

As shown in Figure~\ref{fig:si-detuning-rigidity}, the simultaneous
locking-and-pinning counts at the four adjacent sampled detunings are
$5/5$, $4/5$, $5/5$, and $3/5$, respectively. The median selected
wave vectors remain finite, with $q_0/\pi$ between 0.48 and 0.52 over
these points, and the median laboratory-frame sign errors lie between
0.043 and 0.052. Thus the previously identified pinned period-two
response survives a sampled finite drive-detuning interval. Individual
seeds outside this interval can also meet the criteria, so these data do
not locate either boundary or determine an Arnold-tongue width. In
particular, a wider scan based solely on $\Sigma_g$ locking would also
include rotating registries and cannot by itself be called a pinned DTC
region.
\FloatBarrier

\section{Dimensional mapping and experimental parameters}
\label{sec:si-mapping}

The nondimensionalization also gives a direct illustrative mapping back to
physical units. For chosen particle mass $M$, trap length $\ell$, and
reference frequency $\omega_{\rm ref}$,
\begin{equation}
X_{\rm phys}=\ell x,\qquad
t_{\rm phys}=\frac{t}{\omega_{\rm ref}},\qquad
E_0=M\ell^2\omega_{\rm ref}^2,
\qquad
k_{\rm B}T_{\rm phys}=E_0T.
\label{eq:si-dimensional-conversion}
\end{equation}
Table~\ref{tab:si-experiment} gives an illustrative frequency conversion.
It is an order-of-magnitude feasibility example, not a unique experimental
design. A complete temperature conversion additionally requires the
platform-specific values of $M$ and $\ell$.

\begin{center}
\refstepcounter{table}
\label{tab:si-experiment}
\noindent{\small TABLE~\thetable. \textbf{Illustrative optical-tweezer mapping.}\par}
\vspace{0.6ex}
\begin{ruledtabular}
\begin{tabular}{lll}
Quantity & Dimensionless value & Illustrative dimensional value\\
\hline
Reference frequency & $\omega_0=1$ & $\omega_{\rm ref}/2\pi=27.5$ kHz\\
Damping & $\gamma=0.15$ & $\Gamma/2\pi\simeq4.13$ kHz\\
Drive frequency & $\Omega_D=2.07$ & $\Omega_D^{\rm phys}/2\pi\simeq56.9$ kHz\\
Stiffness modulation & $A_D=0.49$ & $49\%$\\
Mean coupling & $k_0=0.1$ & $K_0=0.1M\omega_{\rm ref}^2$\\
Noise strength & $T$ & $k_{\rm B}T_{\rm phys}=M\ell^2\omega_{\rm ref}^2T$\\
Ring size & $N=100$ & scalable array target\\
\end{tabular}
\end{ruledtabular}
\end{center}

Parametric modulation and stochastic activation have been demonstrated for
levitated particles \cite{ricci2017stochastic}. Reconfigurable arrays with
more than 25 trapped particles have also been realized
\cite{yan2023ondemand,siegel2025arrays}, while tunable nonreciprocal
couplings have been demonstrated in few-particle settings
\cite{rieser2022tunable,reisenbauer2024nonhermitian}. Combining these
capabilities in a stable ring remains an experimental challenge rather than
an established realization of the complete protocol.
\FloatBarrier

\section{Few-site system-size dependence}
\label{sec:si-few-site}

To test whether the four-site result is an isolated exception, we simulated
\[
N\in\{3,4,5,6,8,10,12,16,20,24,32\}
\]
at $(A_D,\nu)=(0.47,0.725),(0.49,0.80),(0.51,0.85)$, with five seeds per
point and $T=10^{-3}$. Each 2400-cycle run retains its final 1200 cycles
at 40 steps/cycle. We select an allowed ring momentum $q_0=2\pi m/N$;
for $N=3$, the sole nonzero pair $q=\pm2\pi/3$ is included explicitly.
Peak interpolation uses fourfold zero padding over
$0.43\leq\omega/\Omega_D\leq0.57$; the true resolution is $1/1200$.

All 165 trajectories retain temporal locking:
$r_\Sigma\geq0.9818$ and
$|\dot\Sigma_g|\leq1.35\times10^{-4}$ rad/cycle. Of these, 150 also have
$|v_\Delta|\geq10^{-3}$ wavelengths/cycle, a half-splitting of at least
$2/1200$ in units of $\Omega_D$, and
$|\epsilon_{\rm split}|\leq0.05$ [Eq.~\eqref{eq:si-residuals}]. Their median
$|\epsilon_{\rm split}|$ is $0.00182$ (maximum $0.0464$), without systematic
growth across the sampled sizes. All spectral midpoints satisfy
$|\omega_c/\Omega_D-1/2|\leq1.05\times10^{-5}$.

\begin{figure}[t]
\centering
\includegraphics[width=0.96\textwidth]{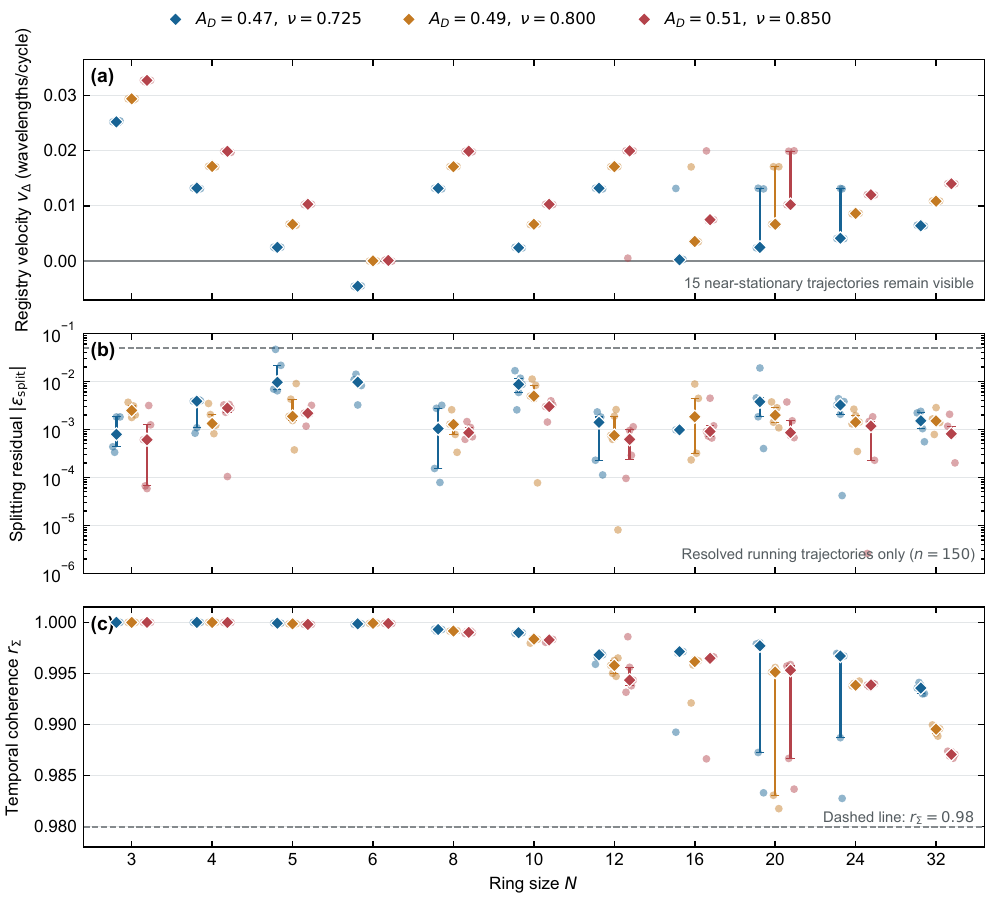}
\caption{\textbf{Few-site system-size dependence of spatial-phase drift.}
(a) Signed phase drift velocity, (b) absolute splitting residual for resolved
rotating realizations, and (c) temporal-phase coherence versus ring size.
Faint symbols show individual seeds; large diamonds and bars show medians and
interquartile ranges for the three indicated parameter points. All 165
trajectories enter (a) and (c); the 15 near-stationary trajectories are
omitted from (b), where unresolved splitting makes the relative residual
ill-conditioned. Dashed lines mark the $|\epsilon_{\rm split}|=0.05$ and
$r_\Sigma=0.98$ diagnostic thresholds. The selected $q_0$ follows the
discrete ring momenta and is listed in Table~\ref{tab:si-few-site}.}
\label{fig:si-few-site}
\end{figure}

The other 15 trajectories are temporally locked but nearly stationary, with
unresolved splitting: ten at $N=6$, one at $N=12$, and four at $N=16$.
Figure~\ref{fig:si-few-site} and Table~\ref{tab:si-few-site} show the full
correspondence at all three points already for $N=3,4,5$ and in rotating
realizations through $N=32$. Thus $N=4$ is not an isolated case; the
rotating-state basin depends on size, parameters, and seed. This scan identifies
neither a minimum ring size nor thermodynamic-limit behavior.

\begin{center}
\refstepcounter{table}
\label{tab:si-few-site}
\noindent{\small TABLE~\thetable. \textbf{Few-site scan by ring size.}
Entries in the three center columns give the number of seeds satisfying
temporal locking, nonzero phase drift velocity, and the resolved splitting
relation, out of five. The last column lists all selected $q_0/\pi$ values
at that size across the three parameter points.\par}
\vspace{0.6ex}
\small
\begin{tabular}{ccccc}
\toprule
$N$ & $(0.47,0.725)$ & $(0.49,0.80)$ & $(0.51,0.85)$ & $q_0/\pi$ \\
\midrule
3  & 5/5 & 5/5 & 5/5 & $2/3$ \\
4  & 5/5 & 5/5 & 5/5 & $1/2$ \\
5  & 5/5 & 5/5 & 5/5 & $2/5$ \\
6  & 5/5 & 0/5 & 0/5 & $1/3$ \\
8  & 5/5 & 5/5 & 5/5 & $1/2$ \\
10 & 5/5 & 5/5 & 5/5 & $2/5$ \\
12 & 5/5 & 5/5 & 4/5 & $1/3,\;1/2$ \\
16 & 1/5 & 5/5 & 5/5 & $3/8,\;1/2$ \\
20 & 5/5 & 5/5 & 5/5 & $2/5,\;1/2$ \\
24 & 5/5 & 5/5 & 5/5 & $5/12,\;1/2$ \\
32 & 5/5 & 5/5 & 5/5 & $7/16$ \\
\bottomrule
\end{tabular}
\end{center}
\FloatBarrier

\end{document}